\documentclass[11pt]{article}

\RequirePackage{fullpage}

\RequirePackage[utf8]{inputenc}

\RequirePackage{amsmath}
\RequirePackage{amssymb}
\RequirePackage{hyperref}

\usepackage{algorithm}
\usepackage[noend]{algpseudocode}

\RequirePackage[amsmath,hyperref,thmmarks]{ntheorem}

\theoremseparator{.}
\newtheorem{definition}{Definition}

\newtheorem{claim}{Claim}
\newtheorem{proposition}{Proposition}
\newtheorem{lemma}{Lemma}
\newtheorem{theorem}{Theorem}
\newtheorem{corollary}{Corollary}

\theorembodyfont{\normalfont}
\newtheorem{remark}{Remark}

\theoremstyle{nonumberplain}
\theoremheaderfont{\normalfont\itshape}
\theoremsymbol{\ensuremath{\Box}}
\newtheorem{proof}{Proof}

\newcommand{\Bit}{\{0,1\}}
\newcommand{\NN}{\mathbb{N}}
\newcommand{\RR}{\mathbb{R}}
\newcommand{\FF}{\mathbb{F}}

\DeclareMathOperator{\E}{\mathbb{E}}
\DeclareMathOperator{\V}{\mathbb{V}}

\newcommand{\op}[1]{\ensuremath{\operatorname{#1}}}

\newcommand{\Aut}{\op{Aut}}
\newcommand{\GL}{\op{GL}}

\newcommand{\poly}{\op{poly}}
\newcommand{\polylog}{\op{polylog}}
\newcommand{\KT}{\op{KT}}

\newcommand{\ceil}[1]{{\left\lceil{#1}\right\rceil}}

\newcommand{\pp}[1]{\ensuremath{\mathrm{#1}}}

\newcommand{\MCSP}{\pp{MCSP}}
\newcommand{\MKTP}{\pp{MKTP}}
\newcommand{\GI}{\pp{GI}}
\newcommand{\GA}{\pp{GA}}
\newcommand{\Iso}{\pp{Iso}}

\newcommand{\RKT}{{R_{\rm KT}}}

\newcommand{\cc}[1]{\ensuremath{\mathsf{#1}}}

\newcommand{\NP}{\cc{NP}}
\newcommand{\coNP}{\cc{coNP}}
\newcommand{\BPP}{\cc{BPP}}
\newcommand{\RP}{\cc{RP}}
\newcommand{\coRP}{\cc{coRP}}
\newcommand{\ZPP}{\cc{ZPP}}
\newcommand{\SZK}{\cc{SZK}}

\newcommand{\ie}{\textrm{i.e.}} 
\newcommand{\eg}{\textrm{e.g.}} 

\title{Minimum Circuit Size, Graph Isomorphism, \\ and Related
  Problems%
\footnote{This work subsumes and significantly strengthens
	the earlier arXiv submission 1511.08189~\cite{agm.arxiv}. See the end of
	Section~\ref{sec:intro} for more details.}
}

\author{%
	Eric Allender%
		\footnote{
			Rutgers University, Piscataway, NJ, USA,
			\tt{allender@cs.rutgers.edu}
		} \and \;
  Joshua A. Grochow%
		\footnote{
			University of Colorado at Boulder, Boulder,
                        CO, USA, \tt{joshua.grochow@colorado.edu}
		} \and \;
	Dieter van Melkebeek%
		\footnote{
			University of Wisconsin--Madison, Madison, WI, USA,
			\tt{dieter@cs.wisc.edu}
		}  \and \;
  Cristopher Moore%
		\footnote{
			Santa Fe Institute, Santa Fe, NM, USA, \tt{moore@santafe.edu}
		} \and \;
  Andrew Morgan%
		\footnote{
			University of Wisconsin--Madison, Madison, WI, USA,
			\tt{amorgan@cs.wisc.edu}
    }
}

\begin{document}

\maketitle

\begin{abstract}
We study the computational power of deciding whether a given truth-table can be
described by a circuit of a given size (the Minimum Circuit Size Problem, or
$\MCSP$ for short), and of the variant denoted as $\MKTP$ where circuit size is
replaced by a polynomially-related Kolmogorov measure. All prior reductions
from supposedly-intractable problems to $\MCSP$ / $\MKTP$ hinged on the
power of $\MCSP$ / $\MKTP$ to distinguish random distributions from
distributions produced by hardness-based pseudorandom generator
constructions. We develop a fundamentally different approach inspired
by the well-known interactive proof system for the complement of Graph
Isomorphism ($\GI$). It yields a randomized reduction with zero-sided
error from $\GI$ to $\MKTP$. We generalize the result and show that
$\GI$ can be replaced by any isomorphism problem for which the
underlying group satisfies some elementary properties. Instantiations
include Linear Code Equivalence, Permutation Group Conjugacy, and
Matrix Subspace Conjugacy. Along the way we develop encodings of
isomorphism classes that are efficiently decodable and achieve
compression that is at or near the information-theoretic optimum;
those encodings may be of independent interest.
\end{abstract}

\section{Introduction}
\label{sec:intro}

Finding a circuit of minimum size that computes a given Boolean function
constitutes the overarching goal in nonuniform complexity theory.
It defines an interesting computational problem in its own right,
the complexity of which depends on the way the Boolean function is specified.
A generic and natural, albeit verbose, way to specify a Boolean function is via
its truth-table. The corresponding decision problem is known as the
Minimum Circuit Size Problem (\MCSP): Given a truth-table and a
threshold $\theta$, does there exist a Boolean circuit of size at most
$\theta$ that computes the Boolean function specified by the
truth-table? The interest in $\MCSP$ dates back to the dawn of
theoretical computer science \cite{trakhtenbrot}. It continues 
today partly due to the fundamental nature of the problem, and partly
because of the work on natural proofs and the connections between
pseudorandomness and computational hardness.

A closely related problem from Kolmogorov complexity theory is the
Minimum KT Problem ($\MKTP$), which deals with compression in the form
of efficient programs instead of circuits. Rather than asking if the
input has a small circuit when interpreted as the truth-table of a
Boolean function, $\MKTP$ asks if the input has a small program that
produces each individual bit of the input quickly. To be more
specific, let us fix a universal Turing machine $U$. We consider 
descriptions of the input string $x$ in the form of a program $d$ such
that, for every bit position $i$, $U$ on input $d$ and $i$ outputs the
$i$-th bit of $x$ in $T$ steps. The $\KT$ cost of such a description is
defined as $|d|+T$, \ie, the bit-length of the program plus the running
time. The $\KT$ complexity of $x$, denoted $\KT(x)$, is the minimum
$\KT$ cost of a description of $x$. $\KT(x)$ is polynomially related
to the circuit complexity of $x$ when viewed as a truth-table (see
Section~\ref{sec:prelim:complexity-measures} for a more formal
treatment). On input a string $x$ and an integer $\theta$, $\MKTP$
asks whether $\KT(x) \leq \theta$. 
 
Both $\MCSP$ and $\MKTP$ are in $\NP$ but are not known to be in
$\cc{P}$ or $\NP$-complete. As such, they are two prominent
candidates for $\NP$-intermediate status. Others include
factoring integers, discrete log over prime fields, graph isomorphism
(\GI), and a number of similar isomorphism problems.

Whereas $\NP$-complete problems all reduce one to another, even
under fairly simple reductions, less is known about the relative
difficulty of presumed $\NP$-intermediate problems. Regarding $\MCSP$
and $\MKTP$, factoring integers and discrete log over prime fields are
known to reduce to both under randomized reductions with zero-sided
error \cite{powerk,rudow}. Recently, Allender and Das \cite{adas} showed
that $\GI$ and all of $\SZK$ (Statistical Zero Knowledge) reduce to
both under randomized reductions with bounded error.

Those reductions and, in fact, \emph{all} prior reductions of
supposedly-intractable problems to $\MCSP$ / $\MKTP$ proceed along
the same well-trodden path. Namely, $\MCSP$ / $\MKTP$ is used as an
efficient statistical test to distinguish random distributions from
pseudorandom distributions, where the pseudorandom distribution arises
from a hardness-based pseudorandom generator construction. In
particular, 
\cite{kab.cai} employs the construction based on the hardness of
factoring Blum integers, 
\cite{powerk,adas,pervasive,rudow} use the construction from \cite{hill} based
on the existence of one-way functions, and 
\cite{powerk,carmosino} make use of the Nisan-Wigderson
construction~\cite{NisanW94}. 
The property that $\MCSP$ / $\MKTP$ breaks the construction implies that the
underlying hardness assumption fails relative to $\MCSP$ / $\MKTP$,
and thus that the supposedly hard problem reduces to $\MCSP$ / $\MKTP$. 

\paragraph{Contributions.}
The main conceptual contribution of our paper is a fundamentally different way
of constructing reductions to $\MKTP$ based on a novel use of known
interactive proof systems. Our approach applies to $\GI$ and a broad class of
isomorphism problems. A common framework for those isomorphism
problems is another conceptual contribution. In terms of results, our
new approach allows us to eliminate the errors in the recent reductions
from $\GI$ to $\MKTP$, and more generally to establish
\emph{zero-sided error} randomized reductions to $\MKTP$ from many 
isomorphism problems within our framework. These include Linear Code
Equivalence, Matrix Subspace Conjugacy, and Permutation Group Conjugacy (see
Section~\ref{sec:iso:corollaries} for the definitions). The technical
contributions mainly consist of encodings of isomorphism classes that
are efficiently decodable and achieve compression that is at or near
the information-theoretic optimum. 

Before describing the underlying ideas, we note that our techniques
remain of interest even in light of the recent quasi-polynomial-time
algorithm for $\GI$ \cite{babai}. For one, $\GI$ is still not known to be
in \cc{P}, and Group Isomorphism stands as a significant obstacle
to this (as stated at the end of~\cite{babai}). Moreover, our
techniques also apply to the other isomorphism problems mentioned
above, for which the current best algorithms are still exponential.

Let us also provide some evidence that our approach for constructing
reductions to $\MKTP$ differs in an 
important way from the existing ones. We claim that the existing
approach can only yield zero-sided error reductions to $\MKTP$ from
problems that are in $\NP \cap \coNP$, a class which $\GI$
and---\textit{a fortiori}---none of the other isomorphism problems
mentioned above are known to reside in. The reason for the claim is
that the underlying hardness assumptions are fundamentally
average-case,%
\footnote{In some settings worst-case to average-case reductions are
  known, but these reductions are themselves randomized with two-sided
  error.
}
which implies that the reduction can have both false positives and
false negatives. For example, in the
papers employing the construction from \cite{hill}, $\MKTP$ is used in
a subroutine to invert a polynomial-time-computable function (see
Lemma~\ref{lemma:mktp-inverts-bbox} in
Section~\ref{sec:prelim:complexity-measures}), and the subroutine may
fail to find an inverse. Given a reliable but imperfect subroutine,
the traditional way to eliminate false positives is to use the subroutine
for constructing an efficiently verifiable membership witness, and only
accept after verifying its validity. As such, the existence of a
traditional reduction without false positives from a language $L$ to
$\MKTP$ implies that $L \in \NP$. Similarly, a traditional reduction
from $L$ to $\MKTP$ without false negatives is only possible if $L \in
\coNP$, and zero-sided error is only possible if $L \in \NP \cap
\coNP$.

\paragraph{Main Idea.}
Instead of using the oracle for $\MKTP$ in the \emph{construction} of a
candidate witness and then verifying the validity of the candidate
without the oracle, we use the power of the oracle in the
\emph{verification} process. This obviates the need for the language
$L$ to be in $\NP \cap \coNP$ in the case of reductions with
zero-sided error.

Let us explain how to implement this idea for $L = \GI$. Recall that
an instance of $\GI$ consists of a pair $(G_0,G_1)$ of graphs on the 
vertex set $[n]$, and the question is whether $G_0 \equiv G_1$, 
\ie, whether there exists a permutation $\pi \in S_n$ such that
$G_1=\pi(G_0)$, where $\pi(G_0)$ denotes the result of applying the
permutation $\pi$ to the vertices of $G_0$. In order to develop a
zero-sided error algorithm for $\GI$, it suffices to develop one
without false negatives. This is because the false positives can
subsequently be eliminated using the known search-to-decision reduction for
$\GI$ \cite{kst}.

The crux for obtaining a reduction without false negatives from $\GI$
to $\MKTP$ is a witness system for the complement $\overline{\GI}$
inspired by the well-known two-round interactive proof system for
$\overline{\GI}$~\cite{GoldreichMW91}. Consider the distribution
$R_G(\pi) \doteq \pi(G)$ where $\pi \in S_n$ is chosen uniformly at
random. By the Orbit--Stabilizer Theorem, for any fixed $G$, $R_G$ is
uniform over a set of size $N \doteq 
n!/|\Aut(G)|$ and thus has entropy $s = \log(N)$,  where
$\Aut(G) \doteq \{ \pi \in S_n \, : \,  \pi(G) = G \}$ denotes the set
of automorphisms of $G$. For ease of exposition, let us assume that
$|\Aut(G_0)|=|\Aut(G_1)|$ (which is actually the hardest case for
$\GI$), so both $R_{G_0}$ and $R_{G_1}$ have the same entropy $s$. 
Consider picking $r \in \Bit$ uniformly at random, and setting $G =
G_r$. If $(G_0,G_1) \in \GI$, the distributions $R_{G_0}$, $R_{G_1}$,
and $R_{G}$ are all identical, and therefore $R_G$ also has entropy
$s$. On the other hand, if $(G_0,G_1) \not\in \GI$, the entropy of
$R_G$ equals $s+1$. The extra bit of information corresponds to the
fact that in the nonisomorphic case each sample of $R_G$ reveals the
value of $r$ that was used, whereas that bit gets lost in the
reduction in the isomorphic case. 

The difference in entropy suggests that a typical sample of $R_G$ can
be compressed more in the isomorphic case than in the nonisomorphic
case. If we can compute some threshold such that $\KT(R_G)$
\emph{never} exceeds the threshold in the isomorphic case, and exceeds 
it with nonnegligible probability in the nonisomorphic case, we have
the witness system for $\overline{\GI}$ that we aimed for: Take a
sample from $R_G$, and use the oracle for $\MKTP$ to check that it
cannot be compressed at or below the threshold. The entropy difference
of 1 may be too small to discern, but we can amplify the difference by
taking multiple samples and concatenating them. Thus, we end up with a
randomized mapping reduction of the following form, where $t$ denotes
the number of samples and $\theta$ the threshold: 
\begin{equation}\label{eq:reduction}
\begin{array}{l}
\text{Pick $r \doteq r_1\ldots r_t \in \Bit^t$ and $\pi_i \in S_n$ for
$i \in [t]$, each uniformly at random.} \\
\text{Output $(y,\theta)$ where $y \doteq y_1\ldots y_t$ and $y_i
  \doteq \pi_i(G_{r_i})$.} 
\end{array}
\end{equation}

We need to analyze how to set the threshold $\theta$ and argue
correctness for a value of $t$ that is polynomially bounded. In order
to do so, let us first consider the case where the graphs $G_0$ and
$G_1$ are \emph{rigid}, \ie, they have no nontrivial automorphisms, or
equivalently, $s = \log(n!)$. 
\begin{itemize}
\item 
	If $G_0 \not\equiv G_1$, the string $y$ contains all of the
        information about the random string $r$ and the $t$ random
        permutations $\pi_1, \ldots, \pi_t$, which amounts to $ts + t
        = t(s+1)$ bits of information.	This implies that $y$ has
        $\KT$-complexity close to $t(s+1)$ with high probability.
\item
	If $G_0 \equiv G_1$, then we can efficiently produce each bit
        of $y$ from the adjacency matrix representation of $G_0$
        ($n^2$ bits) and the function table of permutations $\tau_i
        \in S_n$ (for $i \in [t]$) such that $y_i \doteq \pi_i(G_{r_i}) =
        \tau_i(G_0)$. Moreover, the set of all permutations $S_n$
        allows an efficiently decodable indexing, \ie, there exists an
        efficient algorithm that takes an index $k \in [n!]$ and
        outputs the function table of the $k$-th permutation in $S_n$
        according to some ordering. An example of such an indexing is
        the Lehmer code (see, \eg, \cite[pp. 12-13]{knuth3} for
        specifics). This shows that 
	\begin{equation}\label{eq:KT-bound} 
		\KT(y) \leq t \lceil s \rceil+ (n + \log(t))^c
	\end{equation}
	for some constant $c$, where the first term represents the cost of the $t$
	indices of $\ceil{s}$ bits each,
	and the second term represents the cost of the $n^2$ bits for the adjacency
	matrix of $G_0$ and the polynomial running time of the decoding process. 
\end{itemize}
If we ignore the difference between $s$ and $\ceil{s}$,
the right-hand side of \eqref{eq:KT-bound} becomes $ts+n^c$, which is
closer to $ts$ than to $t(s+1)$ for $t$ any sufficiently large
polynomial in $n$, say $t=n^{c+1}$. Thus, setting $\theta$ halfway
between $ts$ and $t(s+1)$, i.e., $\theta \doteq t(s+\frac{1}{2})$,
ensures that $\KT(y) > \theta$ holds with high probability if $G_0
\not\equiv G_1$, and never holds if $G_0 \equiv G_1$. 
This yields the desired randomized mapping reduction without false
negatives, modulo the rounding issue of $s$ to $\ceil{s}$.
The latter can be handled by aggregating the permutations $\tau_i$ into blocks
so as to make the amortized cost of rounding negligible.
The details are captured in the \hyperref[lemma:blocking]{Blocking Lemma} of
Section~\ref{sec:GI:rigid}.

What changes in the case of {\rm non-rigid} graphs? For ease of
exposition, let us again assume that $|\Aut(G_0)| =
|\Aut(G_1)|$. There are two complications:
\begin{itemize}
\item[(i)]
	We no longer know how to efficiently compute the threshold
        $\theta \doteq t(s+\frac{1}{2})$ because
        $s \doteq \log(N)$ and $N \doteq \log(n!/|\Aut(G_0)|) =
        \log(n!/|\Aut(G_1)|)$ involves the size of the automorphism
        group.
\item[(ii)]
	The Lehmer code no longer provides sufficient compression in the isomorphic
	case as it requires $\log(n!)$ bits per permutation whereas we
        only have $s$ to spend, which could be considerably less than
        $\log(n!)$.
\end{itemize}
In order to resolve (ii) we develop an efficiently decodable indexing
of cosets
for any subgroup of $S_n$ given by a list of generators
(see Lemma~\ref{lemma:graph-coding} in
Section~\ref{sec:GI:quasirigid}). In fact, our scheme even works for
cosets of a subgroup within another subgroup of $S_n$,
a generalization that may be of independent interest (see
Lemma~\ref{lemma:permutation-group-coding} in the Appendix).
Applying our scheme to $\Aut(G)$ and including a minimal list of
generators for $\Aut(G)$ in the description of the program $p$ allows
us to maintain \eqref{eq:KT-bound}. 

Regarding (i), we can deduce a good approximation to the threshold
with high probability by taking, for both choices of $r \in \Bit$,
a polynomial number of samples of $R_{G_r}$ and using the oracle for
$\MKTP$ to compute the exact $\KT$-complexity of their
concatenation. This leads to a randomized reduction from $\GI$ to
$\MKTP$ with bounded error (from which one without false positives
follows as mentioned before), reproving the earlier result of
\cite{adas} using our new approach (see Remark~\ref{remark:GI-BPP} in 
Section~\ref{sec:GI:quasirigid} for more details). 

In order to avoid false negatives, we need to improve the above 
approximation algorithm such that it never produces a value that is
too small, while maintaining efficiency and the property that it
outputs a good approximation with high probability. In order to do so,
it suffices to develop a \emph{probably-correct overestimator} for
the quantity $n!/|\Aut(G)|$, i.e., a randomized algorithm that takes
as input an $n$-vertex graph $G$, produces the correct quantity
with high probability, and never produces a value that is too
small; the algorithm should run in polynomial time with access to an
oracle for $\MKTP$. Equivalently, it suffices to develop a
probably-correct \emph{under}estimator of similar complexity for
$|\Aut(G)|$.  

The latter can be obtained from the known search-to-decision procedures
for $\GI$, and answering the oracle calls to $\GI$ using the above two-sided
error reduction from $\GI$ to $\MKTP$. There are a number of ways to
implement this strategy; here is one that generalizes to a number of
other isomorphism problems including Linear Code Equivalence.
\begin{enumerate}
\item Find a list of permutations that generates a subgroup of $\Aut(G)$
  such that the subgroup equals $\Aut(G)$ with high probability.

  Finding a list of generators for $\Aut(G)$ deterministically reduces to
  $\GI$. Substituting the oracle for $\GI$ by a two-sided error
  algorithm yields a list of permutations that generates $\Aut(G)$ with
  high probability, and always produces a subgroup of $\Aut(G)$. The
  latter property follows from the inner workings of the reduction,
  or can be imposed explicitly by checking every permutation produced
  and dropping it if it does not map $G$ to itself. We use the
  above randomized reduction from $\GI$ to $\MKTP$ as the two-sided
  error algorithm for $\GI$.
\item Compute the order of the subgroup generated by those
  permutations. 

  There is a known deterministic polynomial-time algorithm to do this
  \cite{seress}. 
\end{enumerate}
The resulting probably-correct underestimator for $|\Aut(G)|$ runs in
polynomial time with access to an oracle for $\MKTP$. Plugging it into
our approach, we obtain a randomized reduction from $\GI$ to $\MKTP$
without false negatives. A reduction with zero-sided error follows as
discussed earlier.

Before applying our approach to other isomorphism problems, let us point
out the important role that the Orbit--Stabilizer Theorem
plays. A randomized algorithm for finding generators for a graph's
automorphism group yields a probably-correct underestimator for the
size of the automorphism group, as well as a randomized algorithm for
$\GI$ without false positives. The Orbit--Stabilizer Theorem allows us
to turn a probably-correct underestimator for $|\Aut(G)|$ into a
probably-correct overestimator for the size of the support of $R_G$,
thereby switching the error from one side to the other, and allowing us
to avoid false negatives instead of false positives.

\paragraph{General Framework.}
Our approach extends to several other isomorphism problems.
They can be cast in the following common framework,
parameterized by an underlying family of group actions $(\Omega,H)$
where $H$ is a group that acts on the universe $\Omega$. 
We typically think of the elements of $\Omega$ as abstract objects,
which need to be described in string format in order to be input to a
computer; we let $\omega(z)$ denote the abstract object represented by
the string $z$.
\begin{definition}[Isomorphism Problem]\label{def:iso}
An instance of an Isomorphism Problem consists of a pair $x=(x_0,x_1)$
that determines a universe $\Omega_x$ and a group $H_x$ that acts on
$\Omega_x$ such that $\omega_0(x) \doteq \omega(x_0)$ and
$\omega_1(x) \doteq \omega(x_1)$ belong to $\Omega_x$. Each $h \in
H_x$ is identified with the permutation $h: \Omega_x \to \Omega_x$
induced by the action. The goal is to determine whether there exists
$h \in H_x$ such that $h(\omega_0(x))=\omega_1(x)$.  
\end{definition}
When it causes no confusion, we drop the argument $x$ and simply write
$H$, $\Omega$, $\omega_0$, and $\omega_1$.
We often blur the---sometimes pedantic---distinction between $z$ and
$\omega(z)$. For example, in $\GI$, each $z$ is an $n\times n$
binary matrix (a string of length $n^2$), and represents the abstract
object $\omega(z)$ of a graph with $n$ labeled vertices; thus, in this
case the correspondence between $z$ and $\omega(z)$ is a
bijection. The group $H$ is the symmetric group $S_n$, and the action
is by permuting the labels.

Table~\ref{table:iso} summarizes how the problems we mentioned earlier
can be cast in the framework (see Section~\ref{sec:iso:corollaries}
for details about the last three). 

\begin{table}[h!]
\centering
\begin{tabular}{ |c|c|c| } 
	\hline
	Problem                      & $H$            & $\Omega$
	\\
	\hline
	Graph Isomorphism            & $S_n$          & graphs with $n$ labeled vertices
	\\
	Linear Code Equivalence      & $S_n$          & subspaces of dimension $d$ in $\FF_q^n$
	\\
	Permutation Group Conjugacy  & $S_n$          & subgroups of $S_n$
	\\
	Matrix Subspace Conjugacy           & $\GL_n(\FF_q)$ & subspaces of dimension $d$ in
                                                  $\FF_q^{n \times n}$
	\\
	\hline
\end{tabular}
\caption{Instantiations of the Isomorphism Problem}
\label{table:iso}
\end{table}

We generalize our construction for $\GI$ to any Isomorphism
Problem by replacing $R_G(\pi) \doteq \pi(G)$ where $\pi \in S_n$ is chosen
uniformly at random, by $R_\omega(h) \doteq h(\omega)$ where $h \in H$ is
chosen uniformly at random. The analysis that the construction yields
a randomized reduction without false negatives from the Isomorphism
Problem to $\MKTP$ carries over, provided that the Isomorphism Problem
satisfies the following properties.
\begin{enumerate}
\item
The underlying group $H$ is \emph{efficiently samplable}, and the
action $(\omega,h) \mapsto h(\omega)$ is efficiently computable. We
need this property in order to make sure the reduction is efficient. 

\item
There is an efficiently computable \emph{normal form} for representing 
elements of $\Omega$ as strings. This property trivially holds in the
setting of $\GI$ as there is a unique adjacency matrix that
represents any given graph on the vertex set $[n]$. However,
uniqueness of representation need not hold in
general. Consider, for example, Permutation Group Conjugacy. An
instance of this problem abstractly consists of two permutation groups
$(\Gamma_0,\Gamma_1)$, represented (as usual) by a sequence of
elements of $S_n$ generating each group. In that case there are
many strings representing the same abstract object, \ie, a subgroup
has many different sets of generators.

For the correctness analysis in the isomorphic case it is important
that $H$ acts on the abstract objects, and \emph{not} on the
binary strings that represent them. In particular, the output of the
reduction should only depend on the abstract object $h(\omega)$,
and not on the way $\omega$ was provided as input. This is because the
latter may leak information about the value of the bit $r$ that 
was picked. The desired independence can be guaranteed by applying a
normal form to the representation before outputting the result. In the
case of Permutation Group Conjugacy, this means transforming a set of 
permutations that generate a subgroup $\Gamma$ into a canonical set of
generators for $\Gamma$. 

In fact, it suffices to have an efficiently computable \emph{complete
  invariant} for $\Omega$, \ie, a mapping from representations of
objects from $\Omega$ to strings such that the image only depends on
the abstract object, and is different for different abstract objects.

\item
There exists a probably-correct overestimator for $N \doteq
|H|/|\Aut(\omega)|$ that is computable efficiently with access to an
oracle for $\MKTP$. We need this property to set the threshold
$\theta \doteq t(s+\frac{1}{2})$ with $s \doteq \log(N)$
correctly. 

\item
There exists an encoding for cosets of $\Aut(\omega)$
in $H$ that achieves $\KT$-complexity close to the
information-theoretic optimum (see 
Section~\ref{sec:prelim:encodings} for the definition of an encoding). 
This property ensures that in the isomorphic case the $\KT$-complexity
is never much larger than the entropy.
\end{enumerate}
Properties 1 and 2 are fairly basic. Property 4 may seem to require an
instantiation-dependent approach.  
However, in Section~\ref{sec:flat-coding-lemma}
we develop a \emph{generic} hashing-based encoding scheme
that meets the requirements. In fact, we give a
nearly-optimal encoding scheme for any samplable distribution that is
almost flat, without reference to isomorphism. Unlike the indexings from
Lemma~\ref{lemma:permutation-group-coding} for the special case 
where $H$ is the symmetric group, the generic construction does not
achieve the information-theoretic optimum, but it comes sufficiently
close for our purposes. 

The notion of a probably-correct overestimator in Property 3 can be
further relaxed to that of a \emph{probably-approximately-correct
overestimator}, or \emph{pac overestimator} for short. This is a
randomized algorithm that with high probability outputs a value within
an absolute deviation bound of $\Delta$ from the correct value, and never
produces a value that is more than $\Delta$ below the correct
value. More precisely, it suffices to efficiently compute with access
to an oracle for $\MKTP$ a pac overestimator for $s \doteq
\log(|H|/|\Aut(\omega)|)$ with deviation $\Delta = 1/4$. The
relaxation suffices because of the difference of about 1/2 between the
threshold $\theta$ and the actual $\KT$-values in both the isomorphic
and the non-isomorphic case. As $s = \log|H| - \log|\Aut(\omega)|$, it
suffices to have a pac overestimator for $\log|H|$ and a pac
\emph{under}estimator for $\log|\Aut(\omega)|$, both to within deviation 
$\Delta/2 = 1/8$ and of the required efficiency.

Generalizing our approach for $\GI$, one way to obtain the desired 
underestimator for $\log|\Aut(\omega)|$ is by showing how to
efficiently compute with access to an oracle for $\MKTP$:
\begin{itemize}
\item[(a)] a list $L$ of elements of $H$ that generates a subgroup 
  $\langle L \rangle$ of $\Aut(\omega)$ such that $\langle L \rangle =
  \Aut(\omega)$ with high probability, and 
\item[(b)] a pac underestimator for $\log|\langle L \rangle|$,
  the logarithm of the order of the subgroup generated by a given list
  $L$ of elements of $H$.  
\end{itemize}
Further mimicking our approach for $\GI$, we know how to achieve (a)
when the Isomorphism Problem allows a search-to-decision
reduction. Such a reduction is known for Linear Code Equivalence,
but remains open for problems like Matrix Subspace Conjugacy and
Permutation Group Conjugacy. However, we show that (a) holds for a
\emph{generic} isomorphism problem provided that products and
inverses in $H$ can be computed efficiently (see
Lemma~\ref{lemma:sample-subgroups} in
Section~\ref{sec:iso:conditions}). The proof hinges on the
ability of $\MKTP$ to break the pseudo-random generator construction
of \cite{hill} based on a purported one-way function
(Lemma~\ref{lemma:mktp-inverts-bbox} in 
Section~\ref{sec:prelim:complexity-measures}). 

As for (b), we know how to efficiently compute the order of the
subgroup \emph{exactly} in the case of permutation groups ($H=S_n$),
even without an oracle for $\MKTP$, and in the case of many matrix
groups over finite fields ($H=\GL_n(\FF_q)$) with oracle access to
$\MKTP$, but some cases remain open (see
footnote~\ref{footnote:left-right} in
Section~\ref{sec:iso:conditions} for more details).
Instead, we show how to \emph{generically} construct a \emph{pac
underestimator} with small deviation given access to $\MKTP$ as long as 
products and inverses in $H$ can be computed efficiently, and $H$
allows an efficient complete invariant (see
Lemma~\ref{lemma:pacue-subgroup-order} in 
Section~\ref{sec:iso:conditions}). The first two conditions are
sufficient to efficiently generate a distribution $\widetilde{p}$ on
$\langle L \rangle$ that is uniform to within a small
relative deviation \cite{Bab1991}. The entropy $\widetilde{s}$ of that
distribution equals $\log|\langle L \rangle|$ to within a small
additive deviation. As $\widetilde{p}$ is almost flat, our encoding scheme
from Section~\ref{sec:flat-coding-lemma} shows that $\widetilde{p}$
has an encoding whose length does not exceed $\widetilde{s}$ by much,
and that can be decoded by small circuits. Given an efficient complete
invariant for $H$, we can use an approach similar to the one we used to
approximate the threshold $\theta$ to construct a pac underestimator
for $\widetilde{s}$ with small additive deviation, namely the amortized
$\KT$-complexity of the concatenation of a polynomial number of
samples from $\widetilde{p}$. With access to an oracle for $\MKTP$ we
can efficiently evaluate $\KT$. As a result, we obtain a pac
underestimator for $\log|\langle L \rangle|$ with a small additive
deviation that is efficiently computable with oracle access to $\MKTP$.

The above ingredients allow us to conclude that all
of the isomorphism problems in Table~\ref{table:iso} reduce to
$\MKTP$ under randomized reductions without false negatives. Moreover,
we argue that Properties 1 and 2 are sufficient to generalize the
construction of Allender and Das \cite{adas}, which yields randomized
reductions of the isomorphism problem to $\MKTP$ without false
positives (irrespective of whether a search-to-decision reduction is
known). By combining both reductions, we conclude that all
of the isomorphism problems in Table~\ref{table:iso} reduce to
$\MKTP$ under randomized reductions with zero-sided error. See 
Sections~\ref{sec:iso} and \ref{sec:iso:corollaries} for more details.

\paragraph{Open Problems.}
The difference in compressibility between the isomorphic and
non-isomorphic case is relatively small. As such, our approach is
fairly delicate. Although we believe it yields zero-sided error
reductions to $\MCSP$ as well, we currently do not know whether that
is the case. An open problem in the other direction is to develop
zero-error reductions from all of $\SZK$ to $\MKTP$. We refer to
Section~\ref{sec:conclusion} for further discussion and other future
research directions. 

\paragraph{Relationship with arXiv 1511.08189.}
This report subsumes and significantly strengthens the earlier report
\cite{agm.arxiv}.
\begin{itemize}
\item Whereas \cite{agm.arxiv} only proves the main
  result for $\GI$ on \emph{rigid} graphs, and for Graph
    Automorphism ($\GA$) on arbitrary graphs, this report proves it for 
  $\GI$ on \emph{arbitrary} graphs (which subsumes the result for $\GA$
  on arbitrary graphs). 
\item Whereas \cite{agm.arxiv} only contains the main result for
  $\GI$, this report presents a framework for a
  generic isomorphism problem, and generalizes the main result for
  $\GI$ to any problem within the framework that satisfies some
  elementary conditions. In
  particular, this report shows that the generalization applies to
  Linear Code Equivalence, Permutation Group Conjugacy, and Matrix Subspace
  Conjugacy. The generalization involves the development of a generic
  efficient encoding scheme for samplable almost-flat distributions that is
  close to the information-theoretic optimum, and reductions to $\MKTP$
  for the following two tasks:
  computing a generating set for the automorphism group, 
  and approximating the size of the subgroup generated
  by a given list of elements.
\item The main technical contribution in \cite{agm.arxiv}
  (efficiently indexing the cosets of the automorphism
  group) was hard to follow. This report contains a clean proof
  using a different strategy, which also generalizes to indexing
  cosets of subgroups of any permutation group, answering a question
  that was raised during presentations of \cite{agm.arxiv}. 
\item The exposition is drastically revised.
\end{itemize}

\section{Preliminaries}
\label{sec:prelim}

We assume familiarity with standard complexity theory, including
the bounded-error randomized polynomial-time complexity classes 
$\BPP$ (two-sided error), $\RP$ (one-sided error, \ie, no false
positives), and $\ZPP$ (zero-sided error, \ie, no false positives, no
false negatives, and bounded probability of no output). In the
remainder of this section we provide more details about
$\KT$-complexity, formally define the related notions of
indexing and encoding, and review some background on
graph isomorphism.

\subsection{KT Complexity}
\label{sec:prelim:complexity-measures}

The measure $\KT$ that we informally described in
Section~\ref{sec:intro}, was introduced and formally defined as
follows in \cite{powerk}. We refer to that paper for more background
and motivation for the particular definition.
\begin{definition}[$\KT$]\label{KTdef}
Let $U$ be a universal Turing machine. For each string $x$, define
$\KT_U(x)$ to be 
\begin{align*}
\min \{\, |d| + T : \;\;  (\forall \sigma \in \{0,1,*\}) \; 
(\forall i \leq |x|+1) \; 
U^d(i,\sigma)  \mbox{ accepts in $T$ steps iff $x_i = \sigma$}\,\}.  
\end{align*}
We define $x_i=*$ if $i > |x|$; thus, for $i=|x|+1$ the machine
accepts iff $\sigma=*$. The notation $U^d$ indicates that the machine
$U$ has random access to the description $d$.
\end{definition}
$\KT(x)$ is defined to be equal to $\KT_U(x)$ for a fixed choice of universal
machine $U$ with logarithmic simulation time overhead
\cite[Proposition~5]{powerk}. In particular, if $d$ consists of the
description of a Turing machine $M$ that runs in time $t_M(n)$ and
some auxiliary information $a$ such that $M^a(i) = x_i$ for $i \in
[n]$, then $\KT(x) \leq |a| + c_M T_M(\log n) \log(T_M(\log n))$,
where $n \doteq |x|$ and $c_M$ is a constant depending on $M$. It
follows that 
$(\mu/\log n)^{\Omega(1)} \leq \KT(x) \leq (\mu \cdot \log n)^{O(1)}$
where $\mu$ represents the circuit complexity of the mapping 
$i \mapsto x_i$ \cite[Theorem~11]{powerk}. 

The Minimum KT Problem is defined as 
$\MKTP \doteq \{ (x,\theta) \mid \KT(x) \leq \theta\}$. 
\cite{powerk} showed that an oracle for $\MKTP$ is sufficient to
invert on average any function that can be computed efficiently.
We use the following formulation:
\begin{lemma}[follows from Theorem 45 in
  {\cite{powerk}}] \label{lemma:mktp-inverts-bbox} 
There exists a polynomial-time probabilistic Turing machine $M$ using
oracle access to $\MKTP$ so that the following holds. For any circuit
$C$ on $n$ input bits, 
\[
\Pr\left[ C(M(C, C(\sigma))  = C(\sigma) \right] \geq 1/\poly(n)
\]
where the probability is over the uniform distribution of $\sigma \in
\Bit^n$ and the internal coin flips of $M$.  
\end{lemma}

\subsection{Random Variables, Samplers, Indexings and Encodings} 
\label{sec:prelim:encodings}

A finite probability space consists of a finite sample space $S$,
and a probability distribution $p$ on $S$. Typical sample spaces
include finite groups and finite sets of strings. The probability
distributions underlying our probability spaces are always uniform. 

A \emph{random variable} $R$ is a mapping from the sample space $S$ to 
a set $T$, which typically is the universe $\Omega$ of a group action
or a set of strings. The random variable $R$ with the uniform
distribution on $S$ induces a distribution $p$ on $T$. We sometimes
use $R$ to denote the induced distribution $p$ as well.

The support of a distribution $p$ on a set $T$ is the set of elements
$\tau \in T$ with positive probability $p(\tau)$. A distribution is
\emph{flat} if it is uniform on its support. 
The \emph{entropy} of a distribution $p$ is the expected value of
$\log(1/p(\tau))$.
The \emph{min-entropy} of $p$ is
the largest real $s$ such that $p(\tau) \leq 2^{-s}$ for every $\tau \in T$.
The \emph{max-entropy} of $p$ is
the least real $s$ such that $p(\tau) \geq 2^{-s}$ for every $\tau \in T$.
For a flat distribution, the min-, max-, and ordinary entropy coincide and
equal the logarithm of the size of the support. For two distributions
$p$ and $q$ on the same set $T$, we say that $q$ approximates $p$
within a factor $1+\delta$ if $q(\tau) / (1+\delta) \leq p(\tau) \leq
(1+\delta) \cdot q(\tau)$ for all $\tau \in T$.
In that case, $p$ and $q$ have the same support, and if $p$ has
min-entropy $s$, then $q$ has min-entropy at least $s-\log(1+\delta)$,
and if $p$ has max-entropy $s$, then $q$ has max-entropy at most
$s+\log(1+\delta)$.

A \emph{sampler} within a factor $1+\delta$ for a distribution $p$
on a set $T$ is a random variable $R : \Bit^\ell \to T$ that induces a
distribution that approximates $p$ within a factor $1+\delta$. We say that
$R$ \emph{samples $T$ within a factor $1+\delta$ from length
  $\ell$}. If $\delta=0$ we call the sampler \emph{exact}. The
choice of $\Bit^\ell$ reflects the fact that distributions need to be
generated from a source of random bits. Factors $1+\delta$ with $\delta > 0$
are necessary in order to sample uniform distributions whose support
is not a power of 2. 

We consider ensembles of distributions $\{p_x\}$ where $x$ ranges over
$\Bit^*$.
We call the ensemble \emph{samplable by polynomial-size circuits} if
there exists an ensemble of random variables $\{R_{x,\delta}\}$ where
$\delta$ ranges over the positive rationals such that $R_{x,\delta}$
samples $p_x$ within a factor $1+\delta$ from length $\ell_{x,\delta}$
and $R_{x,\delta}$ can be computed by a circuit of size
$\poly(|x|/\delta)$. We stress that the circuits can depend on the
string $x$, not just on $|x|$. If in addition the mappings  
$(x,\delta) \mapsto \ell_{x,\delta}$ and
$(x,\delta,\sigma) \mapsto R_{x,\delta}(\sigma)$ can be computed in
time $\poly(|x|/\delta)$, we call the ensemble \emph{uniformly
  samplable in polynomial time}.

\medskip

One way to obtain strings with high $\KT$-complexity is as samples
from distributions with high min-entropy.
\begin{proposition}\label{prop:complexity-at-least-entropy}
Let $y$ be sampled from a distribution with min-entropy $s$. For all
$k$, we have $\KT(y) \geq s - k$ except with probability at most
$2^{-k}$.  
\end{proposition}

One way to establish upper bounds on $\KT$-complexity is via
efficiently decodable encodings into integers from a small
range. Encodings with the minimum possible range are referred to as
indexings. We use these notions in various settings. The following
formal definition is for use with random variables and is
general enough to capture all the settings we need. 
It defines an encoding via its decoder $D$; the range of the encoding
corresponds to the domain of $D$. 

\begin{definition}[encoding and indexing]\label{def:encoding}
Let $R: S \to T$ be a random variable. An \emph{encoding} of $R$ is a
mapping $D: [N] \to S$ such that for every $\tau \in T$ there exists
$i \in [N]$ such that $R(D(i)) = \tau$. We refer to $\ceil{\log(N)}$ as the
\emph{length} of the encoding. An \emph{indexing} is an encoding with
$N = |T|$.
\end{definition}

Definition~\ref{def:encoding} applies to a set $S$ by identifying $S$
with the random variable that is the identity mapping on $S$. It applies to the
cosets of a subgroup $\Gamma$ of a group $H$ by considering the random 
variable that maps $h \in H$ to the coset $h\Gamma$. It applies to a
distribution induced by a random variable $R$ by considering the
random variable $R$ itself. 

We say that an ensemble of encodings $\{D_x\}$ is
\emph{decodable by polynomial-size circuits}
if for each $x$ there is a circuit of size $\poly(|x|)$
that computes $D_x(i)$ for every $i \in [N_x]$.
If in addition the mapping $(x,i) \mapsto D_x(i)$ is computable in
time $\poly(|x|)$, we call the ensemble \emph{uniformly decodable in
  polynomial time}.

\subsection{Graph Isomorphism and the Orbit-Stabilizer Theorem} 
\label{sec:prelim:graphs}

Graph Isomorphism ($\GI$) is the computational problem of deciding
whether two graphs, given as input, are isomorphic. 
 A \emph{graph} for us is a simple, undirected graph,
that is, a vertex set $V(G)$, and a set $E(G)$ of unordered pairs of vertices.
An \emph{isomorphism} between two graphs $G_0, G_1$ is a bijection
$\pi\colon V(G_0) \to V(G_1)$ that preserves both edges and non-edges:
$(v,w) \in E(G_0)$ if and only if $(\pi(v), \pi(w)) \in E(G_1)$.
An isomorphism from a graph to itself is an \emph{automorphism};
the automorphisms of a given graph $G$ form a group under
composition, denoted $\Aut(G)$. The Orbit--Stabilizer Theorem implies 
that the number of distinct graphs isomorphic to $G$ equals
$n!/|\Aut(G)|$. A graph $G$ is \emph{rigid} if $|\Aut(G)|=1$,
\ie, the only automorphism is the identity, or equivalently, all $n!$
permutations of $G$ yield distinct graphs. 

More generally, let $H$ be a group acting on a universe $\Omega$.
For $\omega \in \Omega$, each $h\in H$ is an isomorphism from $\omega$
to $h(\omega)$. $\Aut(\omega)$ is the set of isomorphisms from
$\omega$ to itself. By the Orbit--Stabilizer Theorem the number of
distinct isomorphic copies of $\omega$ equals $|H|/|\Aut(\omega)|$.

\section{Graph Isomorphism}
\label{sec:GI}

In this section we show:
\begin{theorem} \label{thm:GI}
	$\GI \in \ZPP^\MKTP$.
\end{theorem}
The crux is the randomized mapping reduction from deciding whether a
given pair of $n$-vertex graphs $(G_0,G_1)$ is in $\GI$ to deciding
whether $(y,\theta) \in \MKTP$, as prescribed by
\eqref{eq:reduction}.  Recall that \eqref{eq:reduction} involves picking
a string $r \doteq r_1\ldots r_t \in \Bit^t$ and permutations $\pi_i$ at
random, and constructing the string
$y = y_1\ldots y_t$, where $y_i = \pi_i(G_{r_i})$.
We show how to determine $\theta$ such 
that a sufficiently large polynomial $t$ guarantees that the reduction
has no false negatives. We follow the outline of
Section~\ref{sec:intro}, take the same four steps, and fill in the
missing details.

\subsection{Rigid Graphs} \label{sec:GI:rigid}

We first consider the simplest setting, in which both $G_0$ and $G_1$
are rigid. We argue that $\theta \doteq  t(s+\frac{1}{2})$ works,
where $s = \log(n!)$. 

\medskip

\noindent
{\it Nonisomorphic Case.} \, If $G_0 \not\equiv G_1$,
then (by rigidity), each choice of $r$ and each distinct sequence of $t$
permutations results in a different string $y$, and thus the distribution
on the strings $y$ has entropy $t(s+1)$ where
$s \doteq \log(n!)$.  Thus, by
Proposition~\ref{prop:complexity-at-least-entropy}, 
$\KT(y) > \theta = t(s+1) - \frac{t}{2}$ with all but exponentially
small probability in $t$. Thus with high probability the algorithm
declares $G_0$ and $G_1$ nonisomorphic. 

\medskip

\noindent
{\it Isomorphic Case.} \, If $G_0 \equiv G_1$, we need to
show that $\KT(y) \leq \theta$ always holds. The key insight is that the
information in $y$ is precisely captured by the $t$ permutations $\tau_1,
\tau_2, \ldots, \tau_t$ such that $\tau_i(G_0) = y_i$. These
permutations exist because $G_0 \equiv G_1$; they are unique by the
rigidity assumption. Thus, $y$ contains at most $ts$ bits of
information. We show that its $\KT$-complexity is not much larger that
this. We rely on the following encoding, due to Lehmer (see, \eg,
\cite[pp. 12--33]{knuth3}):  

\begin{proposition}[Lehmer code] \label{prop:lehmer-coding}
The symmetric groups $S_n$ have indexings that are uniformly decodable
in time $\poly(n)$. 
\end{proposition}

To bound $\KT(y)$, we consider a program $d$ that has the following
information hard-wired into it: $n$, the adjacency matrix of $G_0$,
and the $t$ integers $k_1, \ldots, k_t \in [n!]$
encoding $\tau_1, \ldots, \tau_t$.
We use the decoder from Proposition~\ref{prop:lehmer-coding} to
compute the $i$-th bit of $y$ on input $i$. This can be done in 
time $\poly(n,\log(t))$ given the hard-wired information. 

As mentioned in Section~\ref{sec:intro},
a na\"ive method for encoding the indices $k_1, \ldots, k_t$
only gives the bound $t\ceil{s} + \poly(n,\log(t))$ on $\KT(y)$,
which may exceed $t(s+1)$ and---\emph{a fortiori}---the threshold
$\theta$, no matter how large a polynomial $t$ is. 
We remedy this by aggregating multiple indices into blocks, and 
amortizing the encoding overhead across multiple samples.
The following technical lemma captures the technique. 
For a set $T$ of strings and $b \in \NN$, the statement uses the
notation $T^b$ to denote the set of concatenations of $b$ strings from
$T$; we refer to Section~\ref{sec:prelim:encodings} for the other
terminology. 

\begin{lemma}[Blocking Lemma]\label{lemma:blocking}
Let $\{T_x\}$ be an ensemble of sets of strings such that all strings
in $T_x$ have the same length $\poly(|x|)$.
Suppose that for each $x \in \Bit^*$ and $b \in \NN$,
there is a random variable $R_{x,b}$ whose image contains $(T_x)^b$,
and such that the $R_{x,b}$ is computable by a circuit of size
$\poly(|x|,b)$ and has an encoding of length $s'(x,b)$ decodable by a
circuit of size $\poly(|x|,b)$.
Then there are constants $c_1$ and $c_2$ so that, for every
constant $\alpha > 0$, every $t \in \NN$, every sufficiently
large $x$, and every $y \in (T_x)^t$
\[
\KT(y) \leq t^{1-\alpha}\cdot s'(x, \ceil{t^\alpha}) \;+\;
t^{\alpha\cdot c_1} \cdot |x|^{c_2}.
\]
\end{lemma}
We first show how to apply the \hyperref[lemma:blocking]{Blocking Lemma}
and then prove it.
For a given rigid graph $G$, we let $T_G$ be the image of the random
variable $R_G$ that maps $\pi \in S_n$ to $\pi(G)$ (an adjacency
matrix viewed as a string of $n^2$ bits). We let $R_{G,b}$ be the
$b$-fold Cartesian product of $R_G$, \ie, $R_{G,b}$ takes in $b$
permutations $\tau_1,\ldots,\tau_b \in S_n$, and maps them to 
$\tau_1(G)\tau_2(G)\cdots\tau_b(G)$. $R_{G,b}$ is computable by
(uniform) circuits of size $\poly(n,b)$. To encode an outcome
$\tau_1(G)\tau_2(G)\cdots\tau_b(G)$, we use as index the number whose
base-$(n!)$ representation is written $k_1k_2{\cdots}k_b$, where $k_i$
is the index of $\tau_i$ from the Lehmer code. This indexing has
length $s'(G,b) \doteq \ceil{\log(n!^b)} \leq bs+1$. Given an index,
the list of permutations $\tau_1,\ldots,\tau_b$ can be decoded
by (uniform) circuits of size $\poly(n,b)$. By the
\hyperref[lemma:blocking]{Blocking Lemma}, we have that 
\begin{equation}\label{eq:blocking:calculation}
\KT(y) \leq t^{1-\alpha} (\ceil{t^\alpha} s +1) + 
      t^{\alpha c_1} \cdot n^{c_2}
      \leq
      ts + t^{1-\alpha}\cdot n^{c_0} + t^{\alpha c_1} \cdot n^{c_2}
\end{equation}
for some constants $c_0, c_1, c_2$, every constant $\alpha > 0$, and
all sufficiently large $n$, where we use the fact that $s = \log n! \leq n^{c_0}$.
Setting $\alpha = \alpha_0 \doteq 1/(c_1+1)$, this becomes
$\KT(y) \leq ts + t^{1-\alpha_0} n^{(c_0+c_2)}$.
Taking $t = n^{1+(c_0+c_2)/\alpha_0}$, we see
that for all sufficiently large $n$, $\KT(y) \leq t(s+\frac{1}{2})
\doteq \theta$.

\begin{proof}[of Lemma~\ref{lemma:blocking}]
Let $R_{x,b}$ and $D_{x,b}$ be the hypothesized random variables and
corresponding decoders.
Fix $x$ and $t$, let $m = \poly(|x|)$ denote the length of the strings
in $T_x$, and let $b \in \NN$ be a parameter to be set later.  

To bound $\KT(y)$, we first break $y$ into $\ceil{t/b}$ blocks
$\widetilde{y}_1, \widetilde{y}_2, \ldots, \widetilde{y}_{\ceil{t/b}}$
where each $\widetilde{y}_i \in (T_x)^b$. As the image of $R_{x,b}$
contains $(T_x)^b$, $\widetilde{y}_i$ is encoded by some
index $k_j$ of length $s'(x,b)$. 

Consider a program $d$ that has $x$, $t$, $m$, $b$, the circuit for
computing $R_{x,b}$, the circuit for computing $D_{x,b}$, and the
indices $k_1, k_2, \ldots, k_{\ceil{t/b}}$ hardwired, takes an input
$i \in \NN$, and determines the $i$-th bit of $y$ as follows.
It first computes $j_0, j_1 \in \NN$ so that $i$ points to the $j_1$-th
bit position in $\widetilde{y}_{j_0}$.
Then, using $D_{x,b}$, $k_{j_0}$, $\alpha_{x,b}$, and $j_1$, it finds $\sigma$
such that $R_{x,b}(\sigma)$ equals the $\widetilde{y}_{j_0}$.
Finally, it computes $R_{x,b}(\sigma)$ and outputs the $j_1$-th bit,
which is the $i$-th bit of $y$. 

The bit-length of $d$ is at most $\ceil{t/b} \cdot s'(x,b)$ for the
indices, plus $\poly(|x|,b,\log t)$ for the rest. The time needed by
$p$ is bounded by $\poly(|x|,b,\log t)$. Thus $\KT(y) \leq \ceil{t/b}
\cdot s'(x,b) + \poly(|x|,b,\log t) \leq t/b \cdot s'(x,b) + \poly(|x|,b,\log t)$,
where we used the fact that $s'(x,b) \leq \poly(|x|,b)$. The lemma
follows by choosing $b = \ceil{t^\alpha}$. 
\end{proof}

\subsection{Known Number of Automorphisms}
\label{sec:GI:quasirigid}

We generalize the case of rigid graphs to graphs for which we
know the size of their automorphism groups. 
Specifically, in addition to the two input graphs $G_0$ and $G_1$,
we are also given numbers $N_0, N_1$ where 
$N_i \doteq n!/|\Aut(G_i)|$. Note that if $N_0 \ne N_1$, we can right
away conclude that $G_0 \not\equiv G_1$. Nevertheless, we do not 
assume that $N_0 = N_1$ as the analysis of the case $N_0 \ne N_1$ will
be useful in Section~\ref{sec:GI:assume-pcoe}.

The reduction is the same as in Section~\ref{sec:GI:rigid} with the
correct interpretation of $s$. The main difference lies in the
analysis, where we need to accommodate for the loss in entropy that
comes from having multiple automorphisms. 

Let $s_i \doteq \log(N_i)$ be the entropy in a random permutation of $G_i$.
Set $s \doteq \min(s_0, s_1)$, and $\theta \doteq t(s+\frac{1}{2})$.
In the nonisomorphic case the min-entropy of $y$ is at least $t(s+1)$,
so $\KT(y) > \theta$ with high probability.
In the isomorphic case we upper bound $\KT(y)$ by about $ts$. Unlike
the rigid case, we can no longer afford to encode an entire
permutation for each permuted copy of $G_0$; we need a replacement for
the Lehmer code. The following encoding, applied to $\Gamma = \Aut(G)$,
suffices to complete the argument from Section~\ref{sec:GI:rigid}.

\begin{lemma} \label{lemma:graph-coding}
For every subgroup $\Gamma$ of $S_n$ there exists an indexing of the cosets%
\footnote{The choice of left ($\pi \Gamma$) vs right ($\Gamma \pi$)
  cosets is irrelevant for us; all our results hold for both, and one
  can usually switch from one statement to the other by taking
  inverses. Related to this, there is an ambiguity regarding the order
  of application in the composition $gh$ of two permutations: first
  apply $g$ and then $h$, or vice versa. Both interpretations are
  fine. For concreteness, we assume the
  former. \label{footnote:left-right}}
of $\Gamma$ that is uniformly decodable in polynomial time when
$\Gamma$ is given by a list of generators. 
\end{lemma}
We prove Lemma~\ref{lemma:graph-coding} in the Appendix as a corollary
to a more general lemma that gives, for each $\Gamma \leq H \leq S_n$,
an efficiently computable indexing for the cosets of $\Gamma$ in $H$. 

\begin{remark} \label{remark:GI-BPP}
Before we continue towards Theorem~\ref{thm:GI},
we point out that the above ideas yield an alternate proof that 
$\GI \in \BPP^\MKTP$ (and hence that $\GI \in \RP^\MKTP$). This weaker 
result was already obtained in \cite{adas} along the well-trodden path
discussed in Section~\ref{sec:intro};
this remark shows how to obtain it using our new approach. 

The key observation is that in both the isomorphic and the
nonisomorphic case, with high probability $\KT(y)$ stays away from the 
threshold $\theta$ by a growing margin, Moreover, the above analysis
allows us to efficiently obtain high-confidence approximations of
$\theta$ to within any constant using sampling and queries to the 
$\MKTP$ oracle.  

More specifically, for $i \in \Bit$, let $\widetilde{y}_i$ denote the
concatenation of $\widetilde{t}$ independent samples from $R_{G_i}$. 
Our
analysis shows that $\KT(\widetilde{y}_i) \leq \widetilde{t} s_i +
\widetilde{t}^{1-\alpha_0} n^c$ always holds, and that
$\KT(\widetilde{y}_i) \geq \widetilde{t} s_i -
\widetilde{t}^{1-\alpha_0} n^c$
holds with high probability. Thus, $\widetilde{s}_i \doteq
\KT(\widetilde{y}_i)/\widetilde{t}$ approximates $s_i$ with high confidence to
within an additive deviation of $n^c/\widetilde{t}^{\alpha_0}$. Similarly, 
$\widetilde{s} \doteq \min(\widetilde{s}_0,\widetilde{s}_1)$ approximates $s$ to
within the same deviation margin, and $\widetilde{\theta} \doteq
t(\widetilde{s}+\frac{1}{2})$ approximates $\theta$ to within an additive
deviation of $t n^c/\widetilde{t}^{\alpha_0}$. The latter bound can be made
less than 1 by setting $\widetilde{t}$ to a sufficiently large polynomial
in $n$ and $t$. Moreover, all these estimates can be computed in time 
$\poly(\widetilde{t},n)$ with access to $\MKTP$ as $\MKTP$
enables us to evaluate $\KT$ efficiently. 
\end{remark}

\subsection{Probably-Correct Underestimators for the Number of
  Automorphisms}
\label{sec:GI:assume-pcoe}

The reason the $\BPP^\MKTP$-algorithm in Remark \ref{remark:GI-BPP}
can have false negatives is that the approximation
$\widetilde{\theta}$ to $\theta$ may be too small. Knowing the
quantities $N_i \doteq n!/|\Aut(G_i)|$ exactly allows us to compute
$\theta$ exactly and thereby obviates the possibility of false
negatives. In fact, it suffices to compute overestimates for the
quantities $N_i$ which are correct with non-negligible
probability. We capture this notion formally as follows:

\begin{definition}[probably-correct overestimator] \label{def:estimator}
Let $g: \Omega \to \RR$ be a function, and $M$ a randomized
algorithm that, on input $\omega \in \Omega$, outputs a value
$M(\omega) \in \RR$. We say that $M$ is
a \emph{probably-correct overestimator} for $g$ if, for every $\omega
\in \Omega$, $M(\omega) = g(\omega)$ holds with probability at least
$1/\poly(|\omega|)$, and $M(\omega) > g(\omega)$ otherwise. A
\emph{probably-correct underestimator} for $g$ is defined similarly by 
reversing the inequality.
\end{definition}
We point out that, for any probably-correct over-/underestimator,
taking the min/max among $\poly(|\omega|)$ independent runs yields the
correct value with probability $1 - 2^{-\poly(|\omega|)}$.

We are interested in the case where $g(G) = n!/|\Aut(G)|$. Assuming
this $g$ on a given class of graphs $\Omega$ has a
probably-correct overestimator $M$ computable in randomized 
polynomial time with an $\MKTP$ oracle, we argue that $\GI$ on
$\Omega$ reduces to $\MKTP$ in randomized polynomial time without
false negatives. 

To see this, consider the algorithm that, on input a pair $(G_0,G_1)$
of $n$-vertex graphs, computes $\widetilde{N}_i = M(G_i)$ as estimates
of the true values $N_i = \log(n!/|\Aut(G_i)|)$, and then runs the
algorithm from Section~\ref{sec:GI:quasirigid} using the estimates
$\widetilde{N}_i$. 
\begin{itemize}
\item In the case where $G_0$ and $G_1$ are not isomorphic, if both
  estimates $\widetilde{N}_i$ are correct, then the algorithm detects $G_0
  \not\equiv G_1$ with high probability.
\item In the case where $G_0 \equiv G_1$, if $\widetilde{N}_i = N_i$
we showed in Section~\ref{sec:GI:quasirigid} that the algorithm
always declares $G_0$ and $G_1$ to be isomorphic.
Moreover, increasing $\theta$ can only decrease the probability of
a false negative. As the computed threshold $\theta$ increases as a
function of $\widetilde{N}_i$, and the estimate $\widetilde{N}_i$ is
always at least as large as $N_i$, it follows that $G_0$ and $G_1$ are
always declared isomorphic. 
\end{itemize}

\subsection{Arbitrary Graphs}
\label{sec:GI:pcoe-construction}

A probably-correct overestimator for the function $G \mapsto
n!/|\Aut(G)|$ on \emph{any} graph $G$ can be computed in randomized
polynomial time with access to $\MKTP$. The process is described in full
detail in Section~\ref{sec:intro}, based on a $\BPP^\MKTP$ algorithm for
$\GI$ (taken from Remark~\ref{remark:GI-BPP} or from \cite{adas}). This
means that the setting of Section~\ref{sec:GI:assume-pcoe} is actually
the general one. The only difference is that we no longer obtain a
mapping reduction from $\GI$ to $\MKTP$, but an oracle reduction: We
still make use of \eqref{eq:reduction}, but we need more queries to
$\MKTP$ in order to set the threshold $\theta$.  

This shows that $\GI \in \coRP^\MKTP$. As $\GI \in \RP^\MKTP$ follows
from the known search-to-decision reduction for $\GI$, this concludes
the proof of Theorem~\ref{thm:GI} that $\GI \in \ZPP^\MKTP$.

\section{Estimating the Entropy of Flat Samplable Distributions} 
\label{sec:flat-coding-lemma}

In this section we develop a key ingredient in extending Theorem~\ref{thm:GI}
from $\GI$ to other isomorphism problems that fall within the framework
presented in Section~\ref{sec:intro},
namely efficient near-optimal encodings of cosets of automorphism groups.
More generally, our encoding scheme works well for any samplable
distribution that is flat or almost flat. It allows us to
probably-approximately-correctly underestimate the entropy of such
distributions with the help of an oracle for $\MKTP$.

We first develop our encoding, which only requires the existence of a
sampler from strings of polynomial length. The length of the encoding
is roughly the max-entropy of the distribution, which is the
information-theoretic optimum for flat distributions.
\begin{lemma}[Encoding Lemma] \label{lemma:flat-coding}
Consider an ensemble $\{R_x\}$ of random variables that sample
distributions with max-entropy $s(x)$ from length $\poly(|x|)$.
Each $R_x$ has an encoding of length $s(x) + \log s(x) + O(1)$ that is
decodable by polynomial-size circuits.
\end{lemma}

To see how Lemma~\ref{lemma:flat-coding} performs,
let us apply to the setting of $\GI$.
Consider the random variable $R_G$ mapping a permutation $\pi \in
S_n$ to $\pi(G)$. The induced distribution is flat and has entropy $s =
\log(n!/|\Aut(G)|)$, and each $\pi \in S_n$ can be sampled from strings
of length $O(n\log n)$.
The \hyperref[lemma:flat-coding]{Encoding Lemma} thus yields an encoding of length $s
+ \log s + O(1)$ that is efficiently decodable. 
The bound on the length is worse than Lemma~\ref{lemma:graph-coding}'s
bound of $\ceil{s}$, but will still be sufficient for the generalization of
Theorem~\ref{thm:GI} and yield the result for $\GI$.

We prove the \hyperref[lemma:flat-coding]{Encoding Lemma} using hashing. Here
is the idea. Consider a random hash function $h: \Bit^\ell \to \Bit^m$
where $\ell$ denotes the length of the strings in the domain of $R_x$
for a given $x$, and $m$ is set slightly below $\ell-s$. For any fixed
outcome $y$ of $R_x$, there is a positive constant probability that
no more than about $2^\ell / 2^m \approx 2^s$ of all samples $\sigma \in
\Bit^\ell$ have $h(\sigma)=0^m$, and at least one of these also
satisfies $R_x(\sigma)=y$.
Let us say that \emph{$h$ works for $y$} when both
those conditions hold. In that case---ignoring efficiency
considerations---about $s$ bits of information are sufficient to recover
a sample $\sigma_y$ satisfying $R_x(\sigma_y)=y$ from $h$.

Now a standard probabilistic argument shows that there exists 
a sequence $h_1, h_2, \ldots$ of $O(s)$
hash functions such that for every possible outcome $y$, there is at
least one $h_i$ that works for $y$. Given such a sequence, we can 
encode each outcome $y$ as the index $i$ of a hash function $h_i$ that
works for $y$, and enough bits of information that allow us to
\emph{efficiently} recover $\sigma_y$ given $h_i$. We show that
$s+O(1)$ bits suffice for the standard linear-algebraic family of hash
functions. The resulting encoding has length $s+\log(s)+O(1)$ and is
decodable by circuits of polynomial size.

\begin{proof}[of Lemma~\ref{lemma:flat-coding}]
Recall that a family $\mathcal{H}_{\ell,m}$ of functions from
$\Bit^\ell$ to $\Bit^m$ is \emph{universal} if for any two distinct
$\sigma_0, \sigma_1 \in \Bit^\ell$, the distributions of $h(\sigma_0)$
and $h(\sigma_1)$ for a uniform choice of $h \in \mathcal{H}_{\ell,m}$
are independent and uniform over $\Bit^m$. We make use of the specific
universal family $\mathcal{H}_{\ell,m}^{\mathrm{(lin)}}$ that consists
of all functions of the form $\sigma \mapsto
U\sigma+v$, where $U$ is a binary $(m \times \ell)$-matrix, $v$ is
a binary column vector of dimension $\ell$, and $\sigma$ is also viewed
as a binary column vector of dimension $\ell$
\cite{CarterW79}. Uniformly sampling from
$\mathcal{H}_{\ell,m}^{\mathrm{(lin)}}$ means picking $U$ and $v$
uniformly at random. 

\begin{claim}\label{claim:flat-coding-lemma-subclaim}
Let $\ell,m \in \NN$ and $s \in \RR$. 
\begin{enumerate}
\item
	For every universal family $\mathcal{H}_{\ell,m}$ with
	$m=\ell-\ceil{s}-2$, and for every $S \subseteq \Bit^\ell$ with
  $|S| \geq 2^{\ell-s}$,  
  \[
		\Pr[
			(\exists \sigma \in S) \, h(\sigma)=0^m \textrm{ and }
			|h^{-1}(0^m)| \leq 2^{\ceil{s}+3}
		] \geq \frac{1}{4},
  \]
	where the probability is over a uniformly random choice of
	$h \in \mathcal{H}_{\ell,m}$. 
\item
The sets $h^{-1}(0^m)$ have indexings that are uniformly decodable in
time $\poly(\ell,m)$, where $h$ ranges over
$\mathcal{H}_{\ell,m}^{\mathrm{(lin)}}$.
\end{enumerate}
\end{claim}

Assume for now that the claim holds, and let us continue with the
proof of the lemma.

Fix an input $x$, and let $\ell = \ell(x)$ and $s = s(x)$. Consider
the family $\mathcal{H}_{\ell,m}^{\mathrm{(lin)}}$ with
$m = \ell - \ceil{s} - 2$. 
For each outcome $y$ of $R_x$, let $S_y$ consist of the strings
$\sigma \in \Bit^\ell$ for which $R_x(\sigma) = y$.
Since the distribution induced by $R_x$ has max-entropy $s$, a fraction
at least $1/2^s$ of the strings in the domain of $R_x$ map to $y$.
It follows that $|S_y| \geq 2^{\ell-s}$. 
 
A hash function $h \in \mathcal{H}_{\ell,m}^{\mathrm{(lin)}}$
\emph{works} for 
$y$ if there is some $\sigma \in S_y$ with $h(\sigma) = 0^m$ and
$|h^{-1}(0^m)| \leq 2^{\ceil{s}+3}$. By the first part of
Claim~\ref{claim:flat-coding-lemma-subclaim}, the probability that a
random $h \in \mathcal{H}_{\ell,m}^{\mathrm{(lin)}}$ works for a fixed 
$y$ is at least $1/4$.
If we now pick $3\ceil{s}$ hash functions independently at random, the
probability that none of them work for $y$ is at most
$(3/4)^{3\ceil{s}} < 1/2^s$.
Since there are at most $2^s$ distinct outcomes $y$, a union bound shows
that there exists a sequence of hash functions
$h_1, h_2, \ldots, h_{3\ceil{s}} \in \mathcal{H}_{\ell,m}^{\mathrm{(lin)}}$
such that for every outcome $y$ of $R_x$ there exists
$i_y \in [3\ceil{s}]$ such that $h_{i_y}$ works for $y$. 

The encoding works as follows. Let $D^{\mathrm{(lin)}}$ denote
the uniform decoding algorithm from part 2 of
Claim~\ref{claim:flat-coding-lemma-subclaim} such that
$D^{\mathrm{(lin)}}(h,\cdot)$ decodes the set $h^{-1}(0^m)$.
For each outcome $y$ of $R_x$, let
$j_y \in [2^{\ceil{s}+3}]$ be such that
$D^{\mathrm{(lin)}}(h_{i_y},j_y) = \sigma_y \in S_y$.
Such a $j_y$ exists since $h_{i_y}$ works for $y$. 
Let $k_y = 2^{\ceil{s}+3} i_y + j_y$. Given $h_1,h_2,\ldots, h_{3\ceil{s}}$
and $\ell$ and $m$ as auxiliary information, we can decode $\sigma_y$ from
$k_y$ by parsing out $i_y$ and $j_y$, extracting $h_{i_y}$ from the
auxiliary information, and running $D^{\mathrm{(lin)}}(h_{i_y},j_y)$.
This gives an encoding for $R_x$ of length
$\ceil{s}+3 + \ceil{\log(3\ceil{s})} = s + \log s + O(1)$
that can be decoded in time $\poly(|x|)$ with the hash functions as
auxiliary information. As each hash function can be described using
$(\ell+1)m$ bits and there are $3\ceil{s} \leq \poly(|x|)$ many of
them, the auxiliary information consists of no more than $\poly(|x|)$
bits. Hard-wiring it yields a decoder circuit of size $\poly(|x|)$. 
\end{proof}

For completeness we argue Claim~\ref{claim:flat-coding-lemma-subclaim}.
\begin{proof}[of Claim~\ref{claim:flat-coding-lemma-subclaim}]
For part 1, let $m=\ell-\ceil{s}-2$, and consider the random
variables $X \doteq |h^{-1}(0^m) \cap S|$ and
$Y \doteq |h^{-1}(0^m)|$.
Because of universality we have that $\V(X) \leq \E(X) = |S|/2^m$, and
by the choice of parameters $|S|/2^m  \geq 4$.
By Chebyshev's inequality 
\[ \Pr(X=0) \leq \Pr( |X-\E(X)| \geq \E(X) ) \leq \frac{\V(X)}{(\E(X))^2}
  \leq \frac{1}{\E(X)} \leq \frac{1}{4}. \]
We have that $\E(Y) = 2^\ell/2^m = 2^{\ceil{s}+2}$. By
Markov's inequality
\[ \Pr(Y \geq 2^{\ceil{s}+3}) = \Pr(Y \geq 2 \E(Y)) \leq \frac{1}{2}. \]
A union bound shows that
\[ \Pr(X=0 \text{ or } Y \geq 2^{\ceil{s}+3}) \leq \frac{1}{4} + \frac{1}{2}, \]
from which part 1 follows.

For part 2, note that if $|h^{-1}(0^m)|>0$ then $|h^{-1}(0^m)| =
2^{\ell-r}$ where $r$ denotes the rank of $U$. In that case, 
given $U$ and $v$, we can use 
Gaussian elimination to find binary column vectors $\hat{\sigma}$ and
$\sigma_1, \sigma_2, \ldots, \sigma_{\ell-r}$ such that
$U\hat{\sigma}+v=0^m$ and the $\sigma_i$'s form a basis for the kernel
of $U$. On input $j \in [2^{\ell-r}]$, the 
decoder outputs $\hat{\sigma} + \sum_{i=1}^{\ell-r}j_i \sigma_i$,
where $\sum_{i=1}^{\ell-r} j_i 2^{i-1}$ is the binary expansion of
$j-1$. The image of the decoder is exactly $h^{-1}(0^m)$. 
As the decoding process runs in time $\poly(\ell,m)$ when given $U$
and $u$, this gives the desired indexing. 
\end{proof}

\begin{remark}
The proof of the \hyperref[lemma:flat-coding]{Encoding Lemma} shows a
somewhat more general result:
For any ensemble $\{R_x\}$ of random variables whose domains
consist of strings of length $\poly(|x|)$, and for \emph{any} bound
$s(x)$, the set of outcomes of $R_x$ with probability at least
$1/2^{s(x)}$ has an encoding of length $s(x) + \log s(x) + O(1)$ that
is decodable by a circuit of size $\poly(|x|)$.
In the case of flat distributions of entropy $s(x)$ that
set contains all possible outcomes. 

We also point out that a similar construction (with a single hash
function) was used in \cite{paturi-pudlak} to boost the success
probability of randomized circuits that decide $\cc{CircuitSAT}$
as a function of the number of input variables.%
\footnote{More precisely, suppose there exists a randomized circuit
  family $A$ of size $f(n,m)$ that decides $\cc{CircuitSAT}$ without
  false positives on instances consisting of circuits $C$ with $n$
  input variables and of description length $m$ such that the
  probability of success is at least $1/2^{\alpha n}$. Applying our
  encoding to the set of random bit sequences that make $A$ accept on
  a positive instance $C$, and hard-wiring the input $C$ into the
  circuit $A$, yields an equivalent instance $C'$ on $\alpha n$
  variables of size $f(n,m)+\mu(D)$, where $\mu(D)$ denotes the
  circuit size of $D$. Applying $A$ to the description of this new
  circuit $C'$ yields a randomized circuit $A'$ to decide whether $C$
  is satisfiable without false positives. For the linear-algebraic
  family of hash functions, $A'$ has size $O(f(n,m)
  \polylog(f(n,m)))$. Its success probability is at least
  $1/2^{\alpha^2 n}$, which is larger than $1/2^{\alpha n}$ when
  $\alpha<1$.} 
\end{remark}

In combination with the \hyperref[lemma:blocking]{Blocking Lemma}, the
\hyperref[lemma:flat-coding]{Encoding Lemma} yields upper bounds on
$\KT$-complexity in the case of distributions $p$ that are samplable by
polynomial-size circuits.
More
precisely, if $y$ is the concatenation of $t$ samples from $p$, we can
essentially upper bound the amortized $\KT$-complexity 
$\KT(y)/t$ by the max-entropy of $p$. On the other hand,
Proposition~\ref{prop:complexity-at-least-entropy} shows that
if the samples are picked independently at random, with high
probability $\KT(y)/t$ is not much less than the min-entropy of
$p$. Thus, in the case of flat distributions, $\KT(y)/t$ is a good
\emph{probably-approximately-correct underestimator} for the
entropy, a notion formally defined as follows.
\begin{definition}[probably-approximately-correct underestimator]
Let $g : \Omega \to \RR$ be a function, and $M$ a randomized algorithm
that, on input $\omega \in \Omega$, outputs a value $M(\omega) \in
\RR$. We say that $M$ is a \emph{probably-approximately-correct
  underestimator} (or \emph{pac underestimator}) for $g$ with deviation
$\Delta$ if, for every $\omega \in \Omega$,
$\left|M(\omega) - g(\omega)\right| \leq \Delta$ holds with probability
at least $1/\poly(|\omega|)$, and $M(\omega) < g(\omega)$ otherwise. 
A \emph{probably-approximately-correct overestimator} (or \emph{pac
  overestimator}) for $g$ is defined similarly, by reversing the last 
inequality.
\end{definition}
Similar to the case of probably-correct under-/overestimators, we can boost
the confidence level of a pac under-/overestimator from
$1/\poly(|\omega|)$ to $1 - 2^{-\poly(|\omega|)}$ by taking the
max/min of $\poly(|\omega|)$ independent runs.

More generally, we argue that the amortized $\KT$-complexity of
samples yields a good pac underestimator for the entropy when the
distribution is \emph{almost} flat, \ie, the difference between the
max- and min-entropy is small. As $\KT$ can be evaluated efficiently
with oracle access to $\MKTP$, pac underestimating the entropy of such
distributions reduces to $\MKTP$. 

\begin{corollary}[Entropy Estimator Corollary]\label{cor:flat-KT}
Let $\{p_x\}$ be an ensemble of distributions such that $p_x$ is
supported on strings of the same length $\poly(|x|)$.
Consider a randomized process that on input $x$ computes $\KT(y)/t$,
where $y$ is the concatenation of $t$ independent samples from $p_x$.
If $p_x$ is samplable by circuits of polynomial size,
then for $t$ a sufficiently large polynomial in $|x|$, $\KT(y)/t$ is a
pac underestimator for the entropy of $p_x$ with deviation
$\Delta(x)+o(1)$,
where $\Delta(x)$ is the difference between the min- and max-entropies
of $p_x$.
\end{corollary}

\begin{proof}
	Since the entropy lies between the min- and max-entropies, it suffices
	to show that $\KT(y)/t$ is at least the min-entropy of $p_x$ with high
	probability, and is always at most the max-entropy of $p_x$ 
	(both up to $o(1)$ terms) when $t$ is a sufficiently large polynomial.
	The lower bound follows from
        Proposition~\ref{prop:complexity-at-least-entropy}. It 
	remains to establish the upper bound.

	Let $\{R_{x,\delta}\}$ be the ensemble of random variables witnessing
	the samplability of $\{p_x\}$ by circuits of polynomial size,
	and let $s(x)$ denote the max-entropy of $p_x$.
	The \hyperref[lemma:blocking]{Blocking Lemma} allows us to bound
	$\KT(y)$ by giving an encoding for random variables whose support
	contains the $b$-tuples of samples from $p_x$.
	Let $R'_{x,b}$ denote the $b$-fold Cartesian product of $R_{x,1/b}$.
	$R'_{x,b}$ induces a distribution that approximates to within a factor
	of $(1 + 1/b)^b = O(1)$ the distribution of the $b$-fold Cartesian
	product of $p_x$, which is a distribution of max-entropy $bs(x)$.
	It follows that the distribution induced by $R'_{x,b}$ has min-entropy
	at most $bs(x) + O(1)$. Its support is exactly the $b$-tuples
        of samples from $p_x$.
	Moreover, the ensemble $\{R'_{x,b}\}$ is computable by circuits of
	size $\poly(n,b)$.
	By the \hyperref[lemma:flat-coding]{Encoding Lemma} there exists an
	encoding of $R'_{x,b}$ of length $b s(x) + \log b + \log s(x) + O(1)$
	that is decodable by circuits of polynomial-size.
	The \hyperref[lemma:blocking]{Blocking Lemma} then says that there
	exist constants $c_1$ and $c_2$ so that for all $\alpha > 0$ and
	all sufficiently large $n$
	\begin{align*}
	\KT(y)
	&\leq t^{1-\alpha}\cdot \left(
		\ceil{t^\alpha}\cdot s(x) + \log s(x) + \alpha \log t +
		O(1)
		\right) + t^{\alpha c_1} \cdot n^{c_2} \\
	&\leq t s(x) + t^{1-\alpha}\cdot (n^{c_0} + c_0 \log n + \alpha \log t + O(1)) + t^{\alpha c_1} \cdot n^{c_2}, 
	\end{align*}
	where we use the fact that there exists a constant $c_0$ such that
	$s(x) \leq n^{c_0}$.
	A similar calculation as the one following
	Equation~\eqref{eq:blocking:calculation} shows that 
        $\KT(y) \leq ts(x) + t^{1-\alpha_0}n^{c_0+c_2)}$ for $t \geq
        n^c$ and $n$ sufficiently large, where
	$\alpha_0 = 1/(1+c_1)$ and $c = 1 +
        (1+c_1)(c_0+c_2)$. Dividing both sides by $t$ yields the
        claimed upper bound.
\end{proof}

\section{Generic Isomorphism Problem}
\label{sec:iso}

In Section~\ref{sec:intro} we presented a common framework for
isomorphism problems and listed some instantiations in
Table~\ref{table:iso}.  In this section we state and prove a
generalization of Theorem~\ref{thm:GI} that applies to many problems
in this framework, including the ones from Table~\ref{table:iso}.

\subsection{Generalization}
\label{sec:iso:thm}

The generalized reduction makes use of a complete invariant for the
abstract universe $\Omega$. For future reference, we define the notion
with respect to a representation for an arbitrary ensemble of sets.
\begin{definition}[representation and complete invariant]
Let $\{ \Omega_x \}$ denote an ensemble of sets. A \emph{representation} of
the ensemble is a surjective mapping $\omega: \Bit^* \to \cup_x
\Omega_x$. A \emph{complete invariant} for $\omega$ is a mapping $\nu:
\Bit^* \to \Bit^*$ such that for all strings $x, z_0, z_1$ with 
$\omega(z_0), \omega(z_1) \in \Omega_x$
\[ \omega(z_0) = \omega(z_1) \Leftrightarrow \nu(z_0) = \nu(z_1). \]
\end{definition}
$\omega(z)$ denotes the set element represented by the string $z$. The
surjective property of a representation guarantees that every set
element has at least one string representing it. 

Note that for the function $\nu$ to represent a \emph{normal form}
(rather than just a complete invariant), it would need to be the
case that $\omega(\nu(z)) = \omega(z)$. Although this additional
property holds for all the instantiations we consider, it is not a
requirement. In our setting, all that matters is that $\nu(z)$ only 
depends on the element $\omega(z)$ that $z$ represents, and is
different for different elements.%
\footnote{For complexity-theoretic investigations into the difference
  between complete invariants and normal forms, see, \eg,
\cite{blassGurevich1,blassGurevich2,FortnowGrochowPEq,
  finkelsteinHescott}.}

We are now ready to state the generalization of Theorem~\ref{thm:GI}. 
\begin{theorem}\label{thm:iso}
Let $\pp{Iso}$ denote an Isomorphism Problem as in
Definition~\ref{def:iso}.
Consider the following conditions: 
\begin{enumerate}
\item\label{cond:uniform-sampling} \emph{[action sampler]}
The uniform distribution on $H_x$ is uniformly samplable in
polynomial time, and
the mapping $(\omega,h) \mapsto h(\omega)$ underlying the action 
$(\Omega_x,H_x)$ is computable in $\ZPP$.
\item\label{cond:normal-form} \emph{[complete universe invariant]}
There exists a complete invariant $\nu$ for the representation
$\omega$ that is computable in $\ZPP$.
\item\label{cond:pcoe} \emph{[entropy estimator]}
There exists a probably-approximately-correct overestimator for
$(x,\omega) \mapsto \log\left(|H_x|/|\Aut(\omega)|\right)$ with deviation
$\Delta = 1/4$ that is computable in randomized time $\poly(|x|)$ with
access to an oracle for $\MKTP$. 
\end{enumerate}
With these definitions:
\begin{enumerate}
\item[(a)]
If conditions~\ref{cond:uniform-sampling} and \ref{cond:normal-form}
hold, then $\pp{Iso}\in\RP^\MKTP$.
\item[(b)]
If conditions~\ref{cond:uniform-sampling}, \ref{cond:normal-form}, and
\ref{cond:pcoe} hold, then $\pp{Iso}\in\coRP^\MKTP$.
\end{enumerate}
\end{theorem}
In the case of $\GI$, $\Omega$ denotes the universe of graphs on $n$
vertices (represented as adjacency matrices viewed as strings of
length $n^2$), and $H$ the group of permutations on $[n]$ (represented
as function tables). All conditions in the statement of
Theorem~\ref{thm:iso} are met. The identity mapping can be used
as the complete invariant $\nu$ in condition~\ref{cond:normal-form},
and the probably-correct overestimator for $n!/|\Aut(G)|$ that we
argued in Sections~\ref{sec:intro} and \ref{sec:GI} immediately yields
the pac overestimator for $\log(n!/|\Aut(G)|)$ required in
condition~\ref{cond:pcoe}.

Note that $\log(n!/|\Aut(G)|)$ equals the entropy of the
distribution induced by the random variable $R_G$. In general, the
quantity $\log(|H_x|/|\Aut(\omega)|)$ in condition~\ref{cond:pcoe}
represents the entropy of $\nu(h(\omega))$ when $h \in H_x$ is picked
uniformly at random.  

\begin{proof}[of Theorem~\ref{thm:iso}]
Let $x$ denote an instance of length $n \doteq |x|$, defining a universe
$\Omega$, a group $H$ that acts on $\Omega$, and two elements 
$\omega_i = \omega_i(x)$ for $i \in \Bit$. Both parts (a) and (b) make use of the
random variables $R_i$ for $i \in \Bit$ where $R_i: H \to \Bit^*$
maps $h \in H$ to $\nu(h(\omega_i))$. 

\paragraph{Part (a).}
We follow the approach from \cite{adas}.
Their argument uses Lemma~\ref{lemma:mktp-inverts-bbox},
which states the existence of a randomized polynomial-time machine $M$ with
access to an $\MKTP$ oracle which, given a random sample $y$ from the
distribution induced by a circuit $C$, recovers with non-negligible
probability of success an input $\sigma$ so that $C(\sigma) = y$.
If we can model the $R_i$ as circuits of size $\poly(n)$ that 
take in an element $h$ from $H$ and output $R_i(h)$, this means
that, with non-negligible probability over a random $h_0 \in H$,
$M(R_0, R_0(h_0))$ outputs some $h_1$ so that $h_1(\omega_0) = h_0(\omega_0)$.
The key observation is that when $\omega_0 \equiv \omega_1$,
$R_0$ and $R_1$ induce the same distribution, and therefore, for a
random element $h_0$, $M(R_1, R_0(h_0))$ outputs some $h_1$ so that
$h_1(\omega_0) = h_0(\omega_0)$ with non-negligible probability
probability of success. Thus $\Iso$ can be decided by trying the
above a polynomial number of times, declaring $\omega_0 \equiv
\omega_1$ if a trial succeeds, and declaring $\omega_0 \not\equiv
\omega_1$ otherwise.

We do not know how to model the $R_i$ exactly as circuits of size
$\poly(n)$, but we can do so
approximately. Condition~\ref{cond:uniform-sampling} implies that
we can construct circuits $C_{i,\delta}$ in time $\poly(n/\delta)$
that sample $h(\omega_i)$ within a factor $1+\delta$. Combined with
the $\ZPP$-computability of $\nu$ in
condition~\ref{cond:normal-form} this means that we can construct a 
circuit $C_\nu$ in time $\poly(n)$ such that the composed circuit
$C_\nu \circ C_{i,\delta}$ samples $R_i$ within a factor $1+\delta$
from strings $\sigma$ of length $\poly(n/\delta)$. 
We use the composed circuits in lieu of $R_i$ in the arguments for $M$
above. More precisely, we pick an input $\sigma_0$ for $C_{0,\delta}$
uniformly at random, and compute $\sigma_1 =
M(C_{1,\delta},C_{0,\delta}(\sigma_0))$. Success means that 
$h_1(\omega_0) = h_0(\omega_0)$, where $h_i=C_{i,\delta}(\sigma_i)$. 
The probability of success for an approximation factor of $1+\delta$ is
at least $1/(1+\delta)^2$ times the probability of success in the exact setting,
which is $1/\poly(n/\delta)$ in the isomorphic case. Fixing $\delta$
to any positive constant, a single trial runs in time $\poly(n)$,
success can be determined in $\ZPP$ (by the second part of
condition~\ref{cond:uniform-sampling}), and the probability of success is at
least $1/\poly(n)$ in the isomorphic case. Completing
the argument as in the exact setting above, we conclude that
$\pp{Iso}\in\RP^\MKTP$.

\paragraph{Part (b).}
We generalize the argument from Section~\ref{sec:GI}.
Let $s_i \doteq \log\left( |H|/|\Aut(\omega_i)| \right)$ for
$i \in \Bit$, and let $M$ be the pac overestimator from
condition~\ref{cond:pcoe}. We assume that $M$ has been amplified such
that it outputs a good estimate with probability exponentially close
to 1. Condition~\ref{cond:uniform-sampling} and the
$\ZPP$-computability of $\nu$ imply that the distribution induced by 
$R_i$ is uniformly samplable in polynomial time,
\ie, for each $i \in \Bit$ and $\delta>0$, there is a random variable
$R_{i,\delta}$ that samples $R_i$ within a factor $1+\delta$
from length $\poly(|x|/\delta)$, and that is computable in time
$\poly(|x|/\delta)$.

Let $t \in \NN$ and $\delta$ be parameters to be determined. 
On input $x$, the algorithm begins by computing the estimates
$\widetilde{s}_i = M(x, \omega_i)$ for $i \in \Bit$,
and sets $\widetilde{s} \doteq \min(\widetilde{s}_0,\widetilde{s}_1)$
and $\widetilde{\theta} \doteq t(\widetilde{s}+\frac{1}{2})$. 
The algorithm then samples $r\in\Bit^t$ uniformly, and constructs $y =
(R_{r_i,\delta}(\sigma_i))_{i=1}^t$, where each
$\sigma_i$ is drawn independently and uniformly from
$\Bit^{\poly(n,1/\delta)}$.
If $\KT(y) > \widetilde{\theta}$, the algorithm declares
$\omega_0 \not\equiv \omega_1$;
otherwise, the algorithm declares $\omega_0 \equiv \omega_1$.

\medskip

\noindent
{\it Nonisomorphic Case.} \,
If $\omega_0 \not\equiv \omega_1$,
we need to show $\KT(y) > \widetilde{\theta}$ with high probability.
Since $R_{i,\delta}$ samples $R_i$ within a factor of
$1+\delta$, and $R_i$ is flat with entropy $s_i$,
it follows that $R_{i,\delta}$ has min-entropy at least 
$s_i-\log(1+\delta)$, and that $y$ is sampled from a distribution 
with min-entropy at least 
\[
t(1+\min(s_0,s_1)-\log(1+\delta)).
\]
Since $M$ is a pac overestimator with deviation $\Delta = 1/4$, 
$|\widetilde{s}_0 - s_0| \leq 1/4$ and $|\widetilde{s}_1 - s_1| \leq 1/4$
with high probability.
When this happens, $\widetilde{s} \leq \min(s_0, s_1) + 1/4$,
\[
\widetilde{\theta} \leq t(\min(s_0,s_1) + 3/4),
\]
and Proposition~\ref{prop:complexity-at-least-entropy} guarantees that
$\KT(y) > \widetilde{\theta}$ except with probability exponentially
small in $t$ as long as $\delta$ is a constant such that
$1 - \log(1+\delta) > 3/4$. Such a positive constant $\delta$ exists.

\medskip

\noindent
{\it Isomorphic Case.} \,
If $\omega_0 \equiv \omega_1$, we need to show that $\KT(y) \leq
\widetilde{\theta}$ always holds for $t$ a sufficiently large
polynomial in $n$, and $n$ sufficiently large. 
Recall that, since $\omega_0 \equiv \omega_1$, $R_0$ and $R_1$ induce
the same distribution, so we can view $y$ as the concatenation of $t$
samples from $R_0$. Each $R_0$ is flat, hence has min-entropy equal to
its max-entropy, and the ensemble of all $R_0$ (across all inputs $x$)
is samplable by (uniform) polynomial-size circuits. The 
\hyperref[cor:flat-KT]{Entropy Estimator Corollary} with $\Delta(x)
\equiv 0$ then implies that $\KT(y) \leq t(s_0 + o(1))$ holds whenever
$t$ is a sufficiently large polynomial in $n$, and $n$ is sufficiently
large. In that case, $\KT(y) \leq t(\widetilde{s}+\frac{1}{4}+o(1)) <
\widetilde{\theta}$ holds because $s_0 \leq \widetilde{s} + 1/4$
follows from $M$ being a pac overestimator for $s_0$ with deviation
$1/4$.
\end{proof}

\begin{remark}\label{remark:iso-efficiency}
The notion of efficiency in
conditions~\ref{cond:uniform-sampling},~and~\ref{cond:normal-form}
can be relaxed to mean the underlying algorithm is implementable by a
family of polynomial-size circuits which is constructible in
$\ZPP^\MKTP$.
It is important for our argument that the circuits themselves do not
have oracle access to $\MKTP$, but it is all right for them to be 
constructible in $\ZPP^\MKTP$ rather than $\cc{P}$ or $\ZPP$. 
For example, a sampling procedure that requires knowing the
factorization of some number (dependent on the input $x$) is fine
because the factorization can be computed in $\ZPP^\MKTP$
\cite{powerk} and then can be hard-wired into the circuit.


In particular, this observation yields an alternate way to show that integer
factorization being in $\ZPP^\MKTP$ implies that the discrete log over
prime fields is in $\ZPP^\MKTP$~\cite{rudow}. Recall that an instance
of the discrete log problem consists of a triple $x=(g,z,p)$, where
$g$ and $z$ are integers, and $p$ is a prime, and the goal is to find
an integer $y$ such that $g^y \equiv z \bmod p$, or report that no
such integer exists. The search version is known to reduce to the
decision version in randomized polynomial time, and the above
observation shows that the decision version is in
$\ZPP^\MKTP$. This is because computing the size of the subgroup of
$\FF_p^\times$ generated by $g$ or $z$ reduces to integer factorization, and
can thus be computed in $\ZPP^\MKTP$. 
\end{remark}

\subsection{Construction of Probably-Correct Overestimators}
\label{sec:iso:conditions}

We now discuss some generic methods to satisfy condition~\ref{cond:pcoe}
in Theorem~\ref{thm:iso}, \ie, how to construct a
probably-approximately-correct overestimator for the quantity
$\log(|H|/|\Aut(\omega)|)$ that is computable in $\ZPP^\MKTP$.  

Here is the generalization of the approach we used in
Section~\ref{sec:GI:pcoe-construction} in the context of
$\GI$:
\begin{enumerate}
\item Find a list $L$ of elements of $H$ that generates a subgroup 
  $\langle L \rangle$ of $\Aut(\omega)$ such that $\langle L \rangle =
  \Aut(\omega)$ with high probability.
\item
Pac underestimate $\log|\langle L \rangle|$ with deviation
$1/8$. This yields a pac underestimator for $\log |\Aut(\omega)|$.
\item
Pac overestimate $\log |H|$ with deviation $1/8$.
\item Return the result of step 3 minus the result of step 2. This
  gives a pac overestimator for $\log(|H|/|\Aut(\omega)|)$ with
  deviation $1/4$.
\end{enumerate}
Although in the setting of $\GI$ we used the oracle for $\MKTP$ only
in step 1, we could use it to facilitate steps 2 and 3 as well.

The first step for $\GI$ follows from the known search-to-decision
reduction. It relies on the fact that \emph{Colored Graph Isomorphism}
reduces to $\GI$, where Colored Graph Isomorphism allows one to assign
colors to vertices with the understanding that the isomorphism needs
to preserve the colors. For all of the isomorphism problems in
Table~\ref{table:iso}, finding a set of generators for the
automorphism group reduces to a natural colored 
version of the Isomorphism Problem, but it is not clear whether the
colored version always reduces to the regular version. The latter
reduction is known for Linear Code Equivalence, but remains open for
problems like Permutation Group Conjugacy and Matrix Subspace
Conjugacy.

However, there is a different, \emph{generic} way to achieve step 1
above, namely based on Lemma~\ref{lemma:mktp-inverts-bbox}, \ie, 
the power of $\MKTP$ to efficiently invert on average any efficiently
computable function.

\begin{lemma}\label{lemma:sample-subgroups}
Let $\pp{Iso}$ denote an Isomorphism Problem as in
Definition~\ref{def:iso} that satisfies
conditions~\ref{cond:uniform-sampling} and \ref{cond:normal-form} of 
Theorem~\ref{thm:iso}, and such that products and inverses in $H_x$
are computable in $\BPP^\MKTP$.
There exists a randomized polynomial-time algorithm using oracle access to
$\MKTP$ with the following behavior: 
On input any instance $x$, and any $\omega\in\Omega_x$, the algorithm
outputs a list of generators for a subgroup $\Gamma$ of $\Aut(\omega)$
such that $\Gamma = \Aut(\omega)$ with probability $1 - 2^{-|x|}$. 
\end{lemma}

\begin{proof}
Consider an instance $x$ of length $n \doteq |x|$, and $\omega \in
\Omega_x$. We first argue that the uniform distribution on
$\Aut(\omega)$ is uniformly samplable in polynomial time with oracle
access to $\MKTP$.  Let $R_\omega$ denote the random
variable that maps $h \in H$ to $\nu(h(\omega))$.
As in the proof of Part 1 of Theorem~\ref{thm:iso}, we can
sample $h$ from $H$ uniformly (to within a small constant factor) and
use Lemma~\ref{lemma:mktp-inverts-bbox} to obtain some $h' \in H$
such that $h'(\omega)=h(\omega)$.
In that case, $h^{-1}h'$ is an automorphism of $\omega$.
The key observation is the following:
if $h$ were sampled perfectly uniformly then,
conditioned on success, the distribution of $h^{-1}h'$ is
\emph{uniform} over $\Aut(\omega)$.
Instead, $h$ is sampled uniformly to within a factor $1+\delta$;
in that case $h^{-1}h'$ is uniform on $\Aut(\omega)$ to within a
factor $1+\delta$ and, as argued in the proof of Theorem~\ref{thm:iso},
the probability of success is $1/\poly(n/\delta)$. 

We run the process many times and retain the automorphism $h^{-1}h'$
from the first successful run (if any); $\poly(n/\delta)$ runs
suffice to obtain, with probability $1-2^{-2n}$, an automorphism that
is within a factor $1+\delta$ from uniform over $\Aut(\omega)$. By the
computability parts of conditions~\ref{cond:uniform-sampling} and
\ref{cond:normal-form}, and by the condition that products and
inverses in $H$ can be computed in $\BPP^\MKTP$, each trial runs in
time $\poly(n/\delta)$. Success can be determined in $\ZPP$ as the
group action is computable in $\ZPP$. It follows that the uniform
distribution on $\Aut(\omega)$ is uniformly samplable in polynomial
time with oracle access to $\MKTP$. 

Finally, we argue that a small number of independent samples $h_1,
h_2, \ldots, h_k$ for some constant $\delta>0$ suffice to ensure that
they generate all of $\Aut(\omega)$ with very high probability. Denote
by $\Gamma_i$ the subgroup of $H_x$ generated by
$h_1,\ldots,h_i$. Note that $\Gamma_i$ always is a subgroup of
$\Aut(\omega)$. For $i < k$, if $\Gamma_i$ is not all of
$\Aut(\omega)$, then $|\Gamma_i| \leq |\Aut(\omega)|/2$.
Thus, with probability at least $\frac{1}{2} \cdot
\frac{1}{1+\delta}$, $h_{i+1} \not\in \Gamma_i$, in which case
$|\Gamma_{i+1}| \geq 2 |\Gamma_i|$. For any constant
$\delta>0$, if follows that 
$k \geq \Theta(n + \log|\Aut(\omega)|) = O(\poly(n))$ suffices to
guarantee that $\Gamma_k = \Aut(\omega)$ with probability at least
$1-2^{-2n}$. The lemma follows.
\end{proof}

The second step for $\GI$ followed from the ability to efficiently compute the
order of permutation groups exactly. Efficient exact algorithms
(possibly with access to an oracle for $\MKTP$) are known for larger
classes of groups, including most matrix groups over finite fields,
but not for all.%
\footnote{For many cases where $L \subseteq \GL_n(\FF_q)$,
  \cite{babai2009polynomial} shows how to compute the exact 
  order of $\langle L \rangle$ in $\ZPP$  with oracles for integer
  factorization and the discrete log. Combined with follow-up results of
  \cite{kantorMagaard,liebeckObrien,kantorMagaard2}, the only cases
  that remain open are those over a field of characteristic $2$ where
  $\langle L \rangle$ contains at least one of the Ree groups 
  $^2 F_4(2^{2n+1})$ as a composition factor, and those over a field
  of characteristic $3$ where $\langle L \rangle$ contains at least
  one of the Ree groups $^2 G_2(3^{2n+1})$ as a composition
  factor. The claim follows as integer factorization and discrete log
  can be computed in $\ZPP^\MKTP$.}
We show how to \emph{generically} pac underestimate $\log|\langle L
\rangle|$ with small deviation (step 2), namely under the prior
conditions that only involve $H$, and the additional condition of a
$\ZPP$-computable complete invariant $\zeta$ for $H$.

The construction hinges on the 
\hyperref[cor:flat-KT]{Entropy Estimator Corollary} and viewing
$\log|\langle L \rangle|$ as the entropy of the uniform distribution
$p_L$ on $\langle L \rangle$.
\begin{enumerate}
\item[($\alpha$)] Provided that $p_L$ is samplable by
  circuits of polynomial size, the corollary allows us to pac
  underestimate $\log|\langle L \rangle|$ as $\KT(y)/t$, where $y$ is
  the concatenation of $t$ independent samples from $p_L$.
\item[($\beta$)] If we are able to uniformly sample $\{p_L\}$
  \emph{exactly} in polynomial time (possibly with access to an oracle for
  $\MKTP$), then we can evaluate the estimator $\KT(y)/t$ in
  polynomial time with access to $\MKTP$. This is because the oracle
  for $\MKTP$ lets us evaluate $\KT$ in polynomial time. 
\end{enumerate}
Thus, if we were able to uniformly sample $\{p_L\}$ \emph{exactly} in
polynomial time, we'd be done. We do not know how to do that, but we
can do it \emph{approximately}, which we argue is sufficient. 

The need for a $\ZPP$-computable complete invariant comes in when
representing the abstract group elements as strings. In order to
formally state the requirement, we make the underlying representation
of group elements explicit; we denote it by $\eta$. 

\begin{lemma} \label{lemma:pacue-subgroup-order}
Let $\{ H_x \}$ be an ensemble of groups. Suppose that the ensemble
has a representation $\eta$ such that the uniform distribution on $H_x$
is uniformly samplable in polynomial-time, products and inverses in $H_x$
are computable in $\ZPP$, and there exists a $\ZPP$-computable
complete invariant for $\eta$. Then for any list $L$ of
elements of $H_x$, the logarithm of the order of the group 
generated by $L$, \ie, $\log|\langle L \rangle|$, can be pac
underestimated with any constant deviation $\Delta>0$ in randomized
time $\poly(|x|, |L|)$ with oracle access to $\MKTP$. 
\end{lemma}

\begin{proof}
Let $\zeta$ be the $\ZPP$-computable complete invariant for $\eta$. 
For each list $L$ of elements of $H_x$,
let $p_L$ denote the distribution of $\zeta(h)$ when $h$ is picked
uniformly at random from $\langle L \rangle$. Note that 
$p_L$ is flat with entropy $s = \log|\langle L \rangle|$. 

\begin{claim} \label{claim:pacue-subgroup-order:subclaim}
The ensemble of distributions $\{p_L\}$ is uniformly samplable in
polynomial time. 
\end{claim}
For every constant $\delta>0$, the claim yields a family of random
variables $\{R_{L,\delta}\}$ computable uniformly in
polynomial time such that $R_{L,\delta}$ induces a distribution
$p_{L,\delta}$ that approximates $p_L$ to within a factor
$1+\delta$. Note that the min-entropy of $p_{L,\delta}$ is at least
$s-\log(1+\delta)$, and the max-entropy of $p_{L,\delta}$ at most 
$s+\log(1+\delta)$, thus their difference is no more than
$2\log(1+\delta)$. 

Let $M_\delta(L)$ denote $\KT(y)/t$, where $y$ is the 
concatenation of $t$ independent samples from $p_{L,\delta}$. 
\begin{enumerate}
\item[($\alpha$)]
The \hyperref[cor:flat-KT]{Entropy Estimator Corollary}
guarantees that for any sufficiently large
polynomial $t$, $M_\delta$ is a pac underestimator for the entropy of
$p_{L,\delta}$ with deviation $2\log(1+\delta) + o(1)$, and thus a pac
underestimator for $s = \log|\langle L \rangle|$ with deviation
$3\log(1+\delta)+o(1)$. 
\item[($\beta$)] 
For any polynomial $t$, we can compute $M_\delta$ in polynomial time
with access to an oracle for $\MKTP$. This is because $R_{L,\delta}$
enables us to generate $y$ in polynomial time. We then use the oracle
for $\MKTP$ to compute $\KT(y)$ exactly, and divide by $t$.
\end{enumerate}
Thus, $M_\delta$ meets all the requirements for our estimator as long
as $3\log(1+\delta) < \Delta$, which holds for some positive
constant $\delta$.

This completes the proof of Lemma~\ref{lemma:pacue-subgroup-order}
modulo the proof of the claim.
\end{proof}

The proof of Claim~\ref{claim:pacue-subgroup-order:subclaim} relies on
the notion of \emph{Erd\H{o}s--R\'enyi 
  generators}. A list of generators $L = (h_1,\ldots,h_k)$ is said to be
Erd\H{o}s--R\'enyi with factor $1+\delta$ if a random subproduct of
$L$ approximates the uniform distribution on $\langle L \rangle$
within a factor $1+\delta$, where a random subproduct is obtained by
picking $r_i \in \Bit$ for each $i \in [k]$ uniformly at random, and
outputting $h_1^{r_1}h_2^{r_2}\cdots h_k^{r_k}$.

\begin{proof}[of Claim~\ref{claim:pacue-subgroup-order:subclaim}]
By definition, if $L$ happens to be Erd\H{o}s--R\'enyi with
factor $1+\delta$, then $p_L$ can be sampled to within a factor
$1+\delta$ with fewer than $|L|$ products in $H_x$.

Erd\H{o}s and R\'enyi \cite{erdos.renyi} showed that, for any finite
group $\Gamma$, a list of $\poly(\log|\Gamma|,\log(1/\delta))$ random
elements of $\Gamma$ form an Erd\H{o}s--R\'enyi list of generators with factor
$1+\delta$. For $\Gamma = \langle L \rangle$, this gives a list $L'$ for which
we can sample $p_{L'} = p_L$. By hard-wiring the list $L'$ into the
sampler for $p_{L'}$, it follows that $p_L$ is samplable by circuits
of size $\poly(\log|\langle L\rangle|, \log(1/\delta)) \leq \poly(|L|/\delta)$.

As for \emph{uniformly} sampling $\{p_L\}$ in polynomial time, 
\cite[Theorem 1.1]{Bab1991} gives a randomized algorithm that
generates out of $L$ a list $L'$ of elements from $\langle L \rangle$
that, with probability $1-\varepsilon$, are  Erd\H{o}s--R\'enyi with
factor $1+\delta$. The algorithm runs in time
$\poly(|x|,|L|,\log(1/\delta),\log(1/\varepsilon))$ assuming products and
inverses in $H_x$ can be computed in $\ZPP$. For $\varepsilon =
\delta/|\langle L \rangle|$, the overall distribution of a random
subproduct of $L'$ is within a factor $1+2\delta$ from $p_L$, and
can be generated in time $\poly(|x|,|L|,\log(1/\delta)) \leq
\poly(|x|,|L|,1/\delta)$. As $\delta$ can be an arbitrary positive
constant, it follows that $p_L$ is uniformly samplable in polynomial 
time. 
\end{proof}

Following the four steps listed at the beginning of this section, we
can replace condition~\ref{cond:pcoe} in Theorem~\ref{thm:iso} by
the conditions of Lemma~\ref{lemma:sample-subgroups} (for step 1),
those of Lemma~\ref{lemma:pacue-subgroup-order} (for step 2), and the
existence of an estimator for the size $|H|$ of the sample space as
stated in step 3. This gives the following result:
\begin{theorem} \label{thm:iso-specific}
Let $\pp{Iso}$ denote an Isomorphism Problem as in
Definition~\ref{def:iso}. Suppose that the ensemble $\{ H_x \}$ has a
representation $\eta$ such that conditions~\ref{cond:uniform-sampling}
and \ref{cond:normal-form} of Theorem~\ref{thm:iso} hold as well as the
following additional conditions:
\begin{enumerate}
\setcounter{enumi}{3}
\item\label{cond:basic-operations} \emph{[group operations]}
Products and inverses in $H_x$ are computable in $\ZPP$.
\item\label{cond:cardinality} \emph{[sample space estimator]}
The map $x \mapsto |H_x|$ has a pac overestimator with deviation $\Delta =
1/8$ computable in $\ZPP^\MKTP$.  
\item\label{cond:group-invariant} \emph{[complete group invariant]}
There exists a complete invariant $\zeta$ for the representation
$\eta$ that is computable in $\ZPP$.
\end{enumerate}
Then $\pp{Iso} \in \ZPP^\MKTP$.
\end{theorem}
As was the case for Theorem~\ref{thm:iso}, the conditions of
Theorem~\ref{thm:iso-specific} can be satisfied in a straightforward
way for $\GI$. The representation $\eta$ of the symmetric
groups $S_n$ meets all the requirements that only involve the
underlying group: uniform samplability as in the first part of
condition~\ref{cond:uniform-sampling}, efficient group operations as
in condition~\ref{cond:basic-operations}, the sample space size
$|H|=|S_n|=n!$ can be computed efficiently
(condition~\ref{cond:cardinality}), and the identity mapping
can be used as the complete group invariant $\zeta$
(condition~\ref{cond:group-invariant}). The efficiency of 
the action (the second part of condition~\ref{cond:uniform-sampling})
and condition~\ref{cond:normal-form} about a complete universe
invariant are also met in the same way as before.

We point out that Claim~\ref{claim:pacue-subgroup-order:subclaim} can
be used to show that the uniform distribution on $H_x$ is uniformly
samplable in polynomial time (the first part of
condition~\ref{cond:uniform-sampling}), provided a set of  
generators for $H_x$ can be computed in $\ZPP$. This constitutes
another use of \cite[Theorem 1.1]{Bab1991}.
%
%

On the other hand, the use of \cite[Theorem 1.1]{Bab1991} in the proof
of Theorem~\ref{thm:iso-specific} can be eliminated.
Referring to parts ($\alpha$) and ($\beta$) in the intuition and proof of
Lemma~\ref{lemma:pacue-subgroup-order}, we note the following:
\begin{enumerate}
\item[($\alpha$)] 
The first part of the proof of
Claim~\ref{claim:pacue-subgroup-order:subclaim} relies on 
\cite{erdos.renyi} but not on \cite[Theorem 1.1]{Bab1991}. It shows
that $p_L$ is samplable by polynomial-size circuits, which is
sufficient for the \hyperref[cor:flat-KT]{Entropy Estimator Corollary}
to apply and show that $M_\delta(L)=\KT(y)/t$ is a pac underestimator
for $\log|\langle L \rangle|$ with deviation $3\log(1+\delta)+o(1)$, where
$y$ is the concatenation of $t$ independent samples from
$p_{L,\delta}$ for a sufficiently large polynomial $t$.
\item[($\beta$)]
In the special case where $\langle L \rangle = \Aut(\omega)$, the
first part of the proof of Lemma~\ref{lemma:sample-subgroups} shows
that, for any constant $\delta>0$, 
$p_{L,\delta}$ is \emph{uniformly} samplable in polynomial time with
access to an oracle for $\MKTP$. Once we have generated $y$ with the
help of $\MKTP$, we use $\MKTP$ once more to evaluate $\KT(y)$ and
output $M_\delta(L)=\KT(y)/t$.
\end{enumerate}
This way, for any constant $\delta>0$ we obtain a pac underestimator
$M_\delta$ for $\log|\Aut(\omega)|$ with deviation $3\log(1+\delta)+o(1)$
that is computable in polynomial time with access to
$\MKTP$. 

This alternate construction replaces steps 1 and 2 in the outline from
the beginning of this section. The resulting alternate proof of
Theorem~\ref{thm:iso-specific} is more elementary (as it does not rely
on \cite[Theorem 1.1]{Bab1991}) but does not entirely follow the
approach we used for $\GI$ of first finding a list $L$ of elements
that likely generates $\Aut(\omega)$ (and never generates more) 
and then determining the size of the subgroup generated by $L$.

\begin{remark}
\label{remark:iso-specific-efficiency}
Remark~\ref{remark:iso-efficiency} on relaxing the efficiency
requirement in conditions~\ref{cond:uniform-sampling}
and~\ref{cond:normal-form} of Theorem~\ref{thm:iso} extends similarly
to Theorem~\ref{thm:iso-specific}.
For Theorem~\ref{thm:iso-specific}, it suffices that all the
computations mentioned in conditions~\ref{cond:uniform-sampling},
\ref{cond:normal-form}, \ref{cond:basic-operations}, and
\ref{cond:group-invariant} be do-able by $\ZPP^\MKTP$-constructible
ordinary circuits.
\end{remark}

\section{Instantiations of the Isomorphism Problem}
\label{sec:iso:corollaries}

In this section we argue that Theorem~\ref{thm:iso-specific} applies
to the instantiations of the Isomorphism Problem listed in
Table~\ref{table:iso} (other than $\GI$, which we covered in
Section~\ref{sec:GI}). We describe each problem, provide some
background, and show that the conditions of
Theorem~\ref{thm:iso-specific} hold, thus proving that the problem is
in $\ZPP^\MKTP$. 

\paragraph{Linear code equivalence.} 
A \emph{linear code} over the finite field $\FF_q$ is a $d$-dimensional
linear subspace of $\FF_q^n$ for some $n$.
Two such codes are (permutationally) \emph{equivalent} if there is a
permutation of the $n$ coordinates that makes them equal as subspaces.

\emph{Linear Code Equivalence} is the problem of deciding whether two
linear codes are equivalent, where the codes are specified as the
row-span of a $d \times n$ matrix (of rank $d$), called a
\emph{generator matrix}. 
Note that two different inputs may represent the same code. There
exists a mapping reduction from $\GI$ to Linear Code Equivalence over
any field \cite{pet.roth,grochow.lie};
Linear Code Equivalence is generally thought to be harder than $\GI$.

In order to cast Code Equivalence in our framework,
we consider the family of actions $(S_n,\Omega_{n,d,q})$ where
$\Omega_{n,d,q}$ denotes the linear codes of length $n$ and dimension
$d$ over $\FF_q$, and $S_n$ acts by permuting the coordinates.
To apply Theorem~\ref{thm:iso-specific}, as the underlying group is
$S_n$, we only need to check the efficiency of the action (second part
of condition~\ref{cond:uniform-sampling}) and the complete universe
invariant (condition~\ref{cond:normal-form}). The former holds because
the action only involves swapping columns in the generator matrix. 
For condition~\ref{cond:normal-form} we can define $\nu(z)$ to
be the reduced row echelon form of $z$. This choice works because two
generator matrices define the same code iff they have the same reduced
row echelon form, and it can be computed in polynomial time.
\begin{corollary} \label{cor:code}
	Linear Code Equivalence is in $\ZPP^\MKTP$.
\end{corollary}


\paragraph{Permutation Group Conjugacy.} 
Two permutation groups $\Gamma_0, \Gamma_1 \leq S_n$ are \emph{conjugate} (or
permutationally isomorphic) if there exists a permutation $\pi \in S_n$ such that
$\Gamma_1 = \pi \Gamma_0 \pi^{-1}$;
such a $\pi$ is called a conjugacy.

The \emph{Permutation Group Conjugacy} problem is to decide whether two
subgroups of $S_n$ are conjugate, where the subgroups are specified by a list
of generators.
The problem is known to be in $\NP \cap \cc{coAM}$, and is at least as
hard as Linear Code Equivalence.
Currently the best known algorithm runs in time $2^{O(n)} \poly(|\Gamma_1|)$
\cite{BCQ}---that is, the runtime depends not only on the input size (which is
polynomially related to $n$),
but also on the size of the groups generated by the input permutations,
which can be exponentially larger.  

Casting Permutation Group Conjugacy in the framework is similar to before:
$S_n$ acts on the subgroup by conjugacy. The action is computable in
polynomial time (second part of condition~\ref{cond:uniform-sampling})
as it only involves inverting and composing permutations. It remains
to check condition~\ref{cond:normal-form}.
Note that there are many different lists that generate the same
subgroup. We make use of the normal form provided by the following
lemma.
\newcounter{normal-form}\setcounter{normal-form}{\value{lemma}}
\begin{lemma} \label{lemma:permutation-group-normal-form}
There is a $\poly(n)$-time algorithm $\nu$ that takes as input a list
$L$ of elements of $S_n$, and outputs a list of generators for the
subgroup generated by the elements in $L$ such that for any two lists
$L_0, L_1$ of elements of $S_n$ that generate the same subgroup,
$\nu(L_0) = \nu(L_1)$. 
\end{lemma}
The normal form from Lemma~\ref{lemma:permutation-group-normal-form}
was known to some experts (Babai, personal communication); for
completeness we provide a proof in the Appendix. By
Theorem~\ref{thm:iso-specific} we conclude:
\begin{corollary} \label{cor:perm}
	Permutation Group Conjugacy is in $\ZPP^\MKTP$. 
\end{corollary}

\paragraph{Matrix Subspace Conjugacy.} 
A \emph{linear matrix space} over $\FF_q$ is a $d$-dimensional linear
subspace of $n \times n$ matrices.
Two such spaces $V_0$ and $V_1$ are \emph{conjugate} if there is an invertible
$n \times n$ matrix $X$ such that
$V_1 = X V_0 X^{-1} \doteq \{ X \cdot M \cdot X^{-1} : M \in V_0\}$,
where ``$\cdot$'' represents matrix multiplication.

\emph{Matrix Subspace Conjugacy} is the problem of deciding whether
two linear matrix spaces are conjugate,
where the spaces are specified as the linear span of $d$ linearly independent
$n \times n$ matrices. There exist mapping reductions from $\GI$ and
Linear Code Equivalence to Matrix Subspace Conjugacy \cite{grochow.lie};
Matrix Subspace Conjugacy is generally thought to be harder than Linear Code
Equivalence. 

In order to cast Matrix Subspace Conjugacy in our framework,
we consider the family of actions $(\GL_n(\FF_q),\Omega_{n,d,q})$
where $\GL_n(\FF_q)$ denotes the $n$-by-$n$ general linear group over
$\FF_q$ (consisting of all invertible $n$-by-$n$ matrices over $\FF_q$
with multiplication as the group operation), $\Omega_{n,d,q}$
represents the set of $d$-dimensional subspaces of $\FF_q^{n \times
  n}$, and the action is by conjugation. 
As was the case with Linear Code Equivalence, two inputs may
represent the same linear matrix space, and we use the reduced row
echelon form of $\omega$ when viewed as a matrix in $\FF_q^{d \times
  n^2}$ as the complete universe invariant. This satisfies
condition~\ref{cond:normal-form} of Theorem~\ref{thm:iso-specific}.
The action is computable in polynomial time (second part of
condition~\ref{cond:uniform-sampling}) as it only involves inverting
and multiplying matrices in $\GL_n(\FF_q)$.

The remaining conditions only depend on the underlying group, which is
different from before, namely $\GL_n(\FF_q)$ instead of $S_n$.
Products and inverses in $\GL_n(\FF_q)$ can be computed in polynomial
time (condition~\ref{cond:basic-operations}), and the identity mapping
serves as the complete group invariant
(condition~\ref{cond:group-invariant}). Thus, only the uniform sampler
for $\GL_n(\FF_q)$ (first part of
condition~\ref{cond:uniform-sampling}) and the pac overestimator for
$|\GL_n(\FF_q)|$ (condition~\ref{cond:cardinality}) remain to be
argued.

The standard way of constructing the elements of $\GL_n(\FF_q)$ consists
of $n$ steps, where the $i$-th step picks the $i$-th row as any row
vector that is linearly independent of the $(i-1)$ prior ones.
The number of choices in the $i$-th step is $q^n-q^{i-1}$.
Thus, $|\GL_n(\FF_q)| = \prod_{i=1}^n (q^n-q^{i-1})$ which can be computed
in time $\poly(|x|)$ (condition~\ref{cond:cardinality}). 
It also follows that the probability that a random $(n \times n)$-matrix
over $\FF_q$ is in $\GL_n(\FF_q)$ is at least some positive constant
(independent of $n$ and $q$), which implies that $\{H_x\}$ can be
uniformly sampled in time $\poly(|x|)$, satisfying the first part of
condition~\ref{cond:uniform-sampling}. 

\begin{corollary} 
Matrix Subspace Conjugacy is in $\ZPP^\MKTP$.
\end{corollary}

Before closing, we note that there is an equivalent of the Lehmer code
for $\GL_n(\FF_q)$. We do not need it for our results, but it may be
of interest in other contexts.  
In general, Lehmer's approach works for indexing objects that consist
of multiple components where the set of possible values for the $i$-th
component may depend on the values of the prior components, but the
\emph{number} of possible values for the $i$-th component is
independent of the values of the prior components. An efficiently
decodable indexing follows provided one can efficiently index the
possible values for the $i$-th component given the values of the prior
components. The latter is possible for $\GL_n(\FF_q)$. We include a
proof for completeness.
\begin{proposition} \label{prop:pgl-coding}
For each $n$ and prime power $q$, $\GL_n(\FF_q)$ has an indexing that
is uniformly decodable in time $\poly(n,\log(q))$.  
\end{proposition}
\begin{proof}
Consider the above process. In the $i$-th step, we need to index the 
complement of the subspace spanned by the $i-1$ row vectors
picked thus far, which are linearly independent. This can be done
by extending those $i-1$ row vectors by $n-i+1$ new row vectors to a
full basis, and considering all $q^{i-1}$ linear combinations of the
$i-1$ row vectors already picked, and all $(q^{n-i+1}-1)$
\emph{non-zero} linear combinations of the other basis vectors, and
outputting the sum of the two components. More precisely, on input $k
\in [q^n-q^{i-1}]$, write $k-1$ as $k_0 + k_1 q^{i-1}$ where
$k_0$ and $k_1$ are nonnegative integers with $k_0
< q^{i-1}$, and output $v_0+v_1$ where $v_0$ is the
combination of the $i-1$ row vectors already picked with coefficients
given by the binary expansion of $k_0$, and $v_1$ is linear
combination of the other basis vectors with coefficients given by the
binary expansion of $k_1+1$. Using Gaussian elimination to construct
the other basis vectors, the process runs in time $\poly(n,\log(q))$.
\end{proof}

\section{Future Directions}
\label{sec:conclusion}

We end with a few directions for further research. 

\subsection{What about Minimum Circuit Size?}\label{sec:MCSP}

We suspect that our techniques also apply to $\MCSP$ in place of
$\MKTP$, but we have been unsuccessful in extending them to $\MCSP$ so far. To show our result for
the complexity measure $\mu=\KT$, we showed the following property for
polynomial-time samplable flat distributions $R$: There exists an
efficiently computable bound $\theta(s,t)$ and a polynomial $t$ such
that if $y$ is the concatenation of $t$ independent samples from $R$,
then 
\begin{eqnarray}
\mu(y) & > & \theta(s,t) \textrm{ holds with high probability if $R$ has
  entropy $s+1$, and} \label{eq:no} \\
\mu(y) & \leq & \theta(s,t) \textrm{ always holds if $R$ has entropy
                $s$.} \label{eq:yes}
\end{eqnarray}
We set $\theta(s,t)$ slightly below $\kappa(s+1,t)$ where $\kappa(s,t)
\doteq st$. \eqref{eq:no} followed from a counting argument, and
\eqref{eq:yes} 
by showing that 
\begin{equation}\label{eq:mktp}
\mu(y) \leq \kappa(s,t) \cdot \left(1+\frac{n^c}{t^{\alpha}}\right)
\end{equation}
always holds for some positive constants $c$ and $\alpha$. We
concluded by observing that for a sufficiently large polynomial $t$
the right-hand side of \eqref{eq:mktp} is significantly below 
$\kappa(s+1,t)$. 

Mimicking the approach with $\mu$ denoting circuit complexity,
we set
\[
	\kappa(s,t) = \frac{st}{\log(st)}\cdot\left(1 + (2-o(1))\cdot \frac{\log\log(st)}{\log(st)}\right).
\]
Then \eqref{eq:no} follows from \cite{Yam2011}.
As for \eqref{eq:yes}, the best counterpart to \eqref{eq:mktp} we know of (see,
e.g., \cite{FM2005}) is
\[ \mu(y) \leq 
\frac{st}{\log(st)}\cdot\left(1 + (3+o(1))\cdot \frac{\log\log(st)}{\log(st)}
\right).
\]
However, in order to make the right-hand side of \eqref{eq:mktp} smaller than
$\kappa(s+1,t)$, $t$ needs to be exponential in $s$.

One possible way around the issue is to boost the entropy gap between
the two cases. This would not only show that all our results for
$\MKTP$ apply to $\MCSP$ as well, but could also form the basis for
reductions between different versions of $\MCSP$ (defined in terms of
different circuit models, or in terms of different size parameters), 
and to clarify the relationship between $\MKTP$ and $\MCSP$. Until
now, all of these problems have been viewed as morally equivalent to
each other, although no efficient reduction is known between
\emph{any} two of these, in either direction. Given the central role
that $\MCSP$ occupies, it would be desirable to have a theorem that
indicates that $\MCSP$ is fairly robust to minor changes to its
definition. Currently, this is lacking.

On a related point, it would be good to know how the complexity of
$\MKTP$ compares with the complexity of the $\KT$-random strings: 
$\RKT = \{x : \KT(x) \geq |x|\}$. Until now, all prior reductions from
natural problems to $\MCSP$ or $\MKTP$ carried over to $\RKT$---but
this would seem to require even stronger gap amplification
theorems. The relationship between $\MKTP$ and $\RKT$ is analogous to
the relationship between $\MCSP$ and the special case of $\MCSP$ that
is denoted $\MCSP'$ in \cite{murray.williams}: $\MCSP'$ consists of
truth tables $f$ of $m$-ary Boolean functions that have circuits of
size at most $2^{m/2}$. 

\subsection{Statistical Zero Knowledge} \label{sec:other}

Allender and Das~\cite{adas} generalized their result that
$\GI \in \RP^\MKTP$ to $\SZK \subseteq \BPP^\MKTP$ by applying their
approach to a known $\SZK$-complete problem. Our proof that  
$\GI \in \coRP^\MKTP$ similarly generalizes to $\SZK \subseteq
\BPP^\MKTP$. We use the $\SZK$-complete problem known as Entropy
Difference: Given two circuits $C_0$ and $C_1$ that induce
distributions whose entropy is at least one apart, decide which of the
two has the higher 
entropy \cite{GoldreichV99}. By combining the Flattening Lemma
\cite{GoldreichV99} with the 
\hyperref[cor:flat-KT]{Entropy Estimator Corollary}, one can show that for any
distribution of entropy $s$ sampled by a circuit $C$, the concatenation
of $t$ random samples from $C$ has, with high probability, $\KT$
complexity between $ts - t^{1-\alpha_0}\cdot\poly(|C|)$ and
$ts + t^{1-\alpha_0}\cdot\poly(|C|)$ for some positive constant
$\alpha_0$. Along the lines of Remark~\ref{remark:GI-BPP}, this
allows us to determine which of $C_0$ or $C_1$ has the higher entropy in
$\BPP^\MKTP$.

A natural next question is whether this can be strengthened to show
$\SZK \subseteq \ZPP^\MKTP$. For this it suffices to prove that 
$\SZK$ is in $\RP^\MKTP$ or in $\coRP^\MKTP$ as 
$\SZK$ is closed under complementation \cite{okamoto}.
In fact, the above approach shows that the following variant is in
$\RP^\MKTP$:
Given a circuit $C$ and a threshold $\theta$ with the
promise that $C$ induces a flat distribution of entropy either at
least $\theta+1$ or else at most $\theta-1$, decide whether the former
is the case.
This is the problem Entropy Approximation \cite{GSV1999a} restricted to
flat distributions.
The general version is known to be complete for $\SZK$ under oracle
reductions~\cite[Lemma~5.1]{GSV1999a}, and therefore is in $\ZPP^\MKTP$
if and only if all of $\SZK$ is.
Thus, showing that Entropy Approximation is in $\RP^\MKTP$ is tantamount
to reducing the two-sided error in the known result that
$\SZK \subseteq \BPP^\MKTP$ to zero-sided error. 

This suggests that the difficulty lies in handling non-flat
distributions. For example, it may be the case that the distribution
sampled by $C$ is supported on every string, but the entropy $s$ is
relatively small. In that case, there is no nontrivial worst-case
bound on the $\KT$ complexity of samples from $C$; with positive
probability, $t$ samples from $C$ may have $\KT$-complexity close to
$t$ times the length of each sample, far above $t(s+1)$.

Trying to go beyond $\SZK$, recall that except for the possible use of
the $\MKTP$ oracle in the construction of the probably-correct
overestimator from condition \ref{cond:pcoe} in Theorem~\ref{thm:iso}
(or as discussed in Remark~\ref{remark:iso-efficiency}), 
the reduction in Theorem~\ref{thm:iso} makes only one query to the oracle.
It was observed in \cite{hirahara.watanabe} that the reduction also
works for any relativized $\KT$ problem $\MKTP^A$ 
(where the universal machine for $\KT$ complexity has access to oracle $A$).
More significantly, \cite{hirahara.watanabe} shows that any problem
that is accepted with negligible error probability by a probabilistic
reduction that makes only one query, relative to \emph{every} set
$\MKTP^A$, must lie in $\cc{AM} \cap \cc{coAM}$. Thus, without
significant modification, our techniques cannot be used in order 
to reduce any class larger than $\cc{AM} \cap \cc{coAM}$ to $\MKTP$.

The property that only one query is made to the oracle was
subsequently used in order to show that $\MKTP$ is hard for the
complexity class $\cc{DET}$ under mapping reductions computable in
nonuniform $\cc{NC}^0$ \cite{allender.hirahara}. 
Similar hardness results (but for a more powerful class of reducibilities) hold
also for $\MCSP$ \cite{oliveira.santhanam}.
This has led to unconditional lower bounds on the circuit complexity of $\MKTP$
\cite{allender.hirahara,hirahara.santhanam},
showing that $\MKTP$ does not lie in the complexity class $\cc{AC}^0[p]$ for any
prime $p$;
it is still open whether similar circuit lower bounds hold for $\MCSP$.

\subparagraph*{Acknowledgments.}
EA acknowledges the support of National Science Foundation grant
CCF-1555409.
JAG was supported by an Omidyar Fellowship from the Santa Fe Institute and
National Science Foundation grant DMS-1620484.
DvM and AM acknowledge the support of National Science Foundation
grant CCF-1319822.
We thank
V.~Arvind for helpful comments about the graph automorphism problem and rigid
graphs,
Alex Russell and Yoav Kallus for helpful ideas on encoding and decoding graphs,
Laci Babai and Peter Brooksbank for answering questions about
computational group theory,
and Oded Goldreich and Salil Vadhan for answering questions about $\SZK$.

\bibliographystyle{alpha}
\bibliography{refer}

\appendix

\section*{Appendix: Coset Indexings and Normal Forms for Permutation
  Groups} 
\label{sec:permutation-group-coding}

In this appendix we develop the efficiently decodable indexings for
cosets of permutation subgroups claimed in
Lemma~\ref{lemma:graph-coding}, and also use some of the underlying
ideas to establish the normal form for permutation groups stated in
Lemma~\ref{lemma:permutation-group-normal-form}.

\paragraph{Indexing Cosets.}
The indexings are not strictly needed for our main results as the
generic encoding from the \hyperref[lemma:flat-coding]{Encoding Lemma}
can be used as a substitute.
However, the information-theoretic optimality of the
indexings may be useful in other contexts. In fact, we present a
further generalization that may be of independent interest, namely an
efficiently decodable indexing for cosets of permutation subgroups
within another permutation subgroup.

\begin{lemma}\label{lemma:permutation-group-coding}
For all $\Gamma \leq H \leq S_n$, there exists an indexing of the
cosets%
\footnote{Recall footnote~\ref{footnote:left-right} on
  page~\pageref{footnote:left-right}.} 
of $\Gamma$ within $H$ that is uniformly decodable in polynomial time
when $\Gamma$ and $H$ are given by a list of generators.
\end{lemma}

Lemma~\ref{lemma:graph-coding} is just the instantiation of
Lemma~\ref{lemma:permutation-group-coding} with $H = S_n$.
The proof of Lemma~\ref{lemma:permutation-group-coding} requires some
elements of the theory of permutation groups.
Given a list of permutations $\pi_1, \ldots, \pi_k \in S_n$, we write
$\Gamma = \langle \pi_1, \ldots, \pi_k \rangle \leq S_n$
for the subgroup they generate.
Given a permutation group $\Gamma \leq S_n$ and a point $i \in [n]$,
the $\Gamma$-orbit of $i$ is the set $\{g(i) : g \in \Gamma\}$,
and the $\Gamma$-stabilizer of $i$ is the subgroup
$\{g \in \Gamma : g(i)=i\} \leq \Gamma$.

We make use of the fact that 
(a) the number of cosets of a subgroup $\Gamma$ of a group $H$ equals
  $|H|/|\Gamma|$, and
(b) the orbits of a subgroup $\Gamma$ of $H$ form a refinement of the
  orbits of $H$.
We also need the following basic routines from computational group
theory (see, for example, \cite{Holt05,seress}).
\begin{proposition}\label{prop:cgt}
Given a set of permutations that generate a subgroup $\Gamma \leq S_n$,
the following can be computed in time polynomial in $n$:
\begin{itemize}
\item[(1)] the cardinality $|\Gamma|$,
\item[(2)] a permutation in $\Gamma$ that maps $u$ to $v$ for given $u,v
  \in [n]$, or report that no such permutation exists in $\Gamma$, and
\item[(3)] a list of generators for the subgroup $\Gamma_v$ of $\Gamma$ that
  stabilizes a given element $v \in [n]$.
\end{itemize}
\end{proposition}

The proof of Lemma~\ref{lemma:permutation-group-coding} makes implicit
use of an efficient process for finding a \emph{canonical representative}
of $\pi \Gamma$ for a given permutation $\pi \in H$, where ``canonical''
means that the representative depends on the coset $\pi \Gamma$
only. The particular canonical representative the process produces can
be specified as follows.
\begin{definition}\label{def:canonical}
For a permutation $\pi \in S_n$ and a subgroup $\Gamma \leq S_n$, the
\emph{canonical representative} of $\pi$ modulo $\Gamma$, denoted $\pi
\bmod \Gamma$, is the lexicographically least $\pi' \in \pi \Gamma$,
where the lexicographic ordering is taken by viewing a permutation
$\pi'$ as the sequence $(\pi'(1), \pi'(2), \dotsc, \pi'(n))$.
\end{definition}

We describe the process as it provides intuition for the proof of
Lemma~\ref{lemma:permutation-group-coding}.

\begin{lemma} \label{lemma:canonical}
There exists a polynomial-time algorithm that takes as input
a generating set for a subgroup $\Gamma \leq S_n$ and a permutation
$\pi \in S_n$, and outputs the canonical representative $\pi \bmod
\Gamma$. 
\end{lemma}

\begin{proof}[of Lemma \ref{lemma:canonical}]
Consider the element 1 of $[n]$. Permutations in $\pi \Gamma$ map 1 to an
element $v$ in the same $\Gamma$-orbit as $\pi(1)$, and for every element
$v$ in the $\Gamma$-orbit of $\pi(1)$ there exists a permutation in $\pi \Gamma$
that maps 1 to $v$. We can canonize the behavior of $\pi$ on the
element 1 by replacing $\pi$ with a permutation $\pi_1 \in \pi \Gamma$ that 
maps 1 to the minimum element $m$ in the $\Gamma$-orbit of $\pi(1)$. This
can be achieved by multiplying $\pi$ to the right with a permutation
in $\Gamma$ that maps $\pi(1)$ to $m$.

Next we apply the same process to $\pi_1$ but consider the
behavior on the element 2 of $[n]$. Since we are no longer allowed to
change the value of $\pi_1(1)$, which equals $m$, the canonization of
the behavior on 2 can only use multiplication on the right with
permutations in $\Gamma_m$, i.e., permutations in $\Gamma$ that stabilize the
element $m$. Doing so results in a permutation $\pi_2 \in \pi_1 \Gamma$.

We repeat this process for all elements $k \in [n]$ in order. In the
$k$-th step, we canonize the behavior on the element $k$ by multiplying
on the right with permutations in $\Gamma_{\pi_{k-1}([k-1])}$, i.e.,
permutations in $\Gamma$ that pointwise stabilize all of the elements
$\pi_{k-1}(\ell)$ for $\ell \in [k-1]$. 
\end{proof}

\begin{proof}[of Lemma \ref{lemma:permutation-group-coding}]
The number of canonical representatives modulo $\Gamma$ in $H$ equals the
number of distinct (left) cosets of $\Gamma$ in $H$, which is
$|H|/|\Gamma|$. We construct an 
algorithm that takes as input a list of generators for $\Gamma$ and
$H$, and an index $i \in [|H|/|\Gamma|]$, 
and outputs the permutation $\sigma$ that is the lexicographically
$i$-th canonical representative modulo $\Gamma$ in $H$.

The algorithm uses a prefix search to construct $\sigma$. In the $k$-th 
step, it knows the prefix $(\sigma(1),\sigma(2),\ldots,\sigma(k-1))$
of length $k-1$, and needs to figure out the correct value $v \in [n]$
to extend the prefix with. In order to do so, the algorithm needs to
compute for each $v \in [n]$ the count $c_v$ of canonical
representatives modulo $\Gamma$ in $H$ that agree with $\sigma$ on $[k-1]$
and take the value $v$ at $k$. The following claims allow us to do
that efficiently when given a permutation $\sigma_{k-1} \in H$ that
agrees with $\sigma$ on $[k-1]$. The claims use the notation $T_{k-1}
\doteq \sigma_{k-1}([k-1])$, which also equals $\sigma([k-1])$.

\begin{claim}\label{claim:1}
The canonical representatives modulo $\Gamma$ in $H$ that agree with
$\sigma \in H$ on $[k-1]$ are exactly the canonical representatives modulo 
$\Gamma_{T_{k-1}}$ in $\sigma_{k-1}H_{T_{k-1}}$.
\end{claim}

\begin{proof}
The following two observations imply Claim \ref{claim:1}.
\begin{itemize}
\item[(i)]
A permutation $\pi \in H$ agrees with $\sigma \in H$ on $[k-1]$ \\
$\Leftrightarrow$ $\pi$ agrees with $\sigma_{k-1}$ on $[k-1]$ \\
$\Leftrightarrow$ $\sigma_{k-1}^{-1} \pi \in H_{T_{k-1}}$ \\
$\Leftrightarrow$ $\pi \in \sigma_{k-1} H_{T_{k-1}}$. 
\item[(ii)]
Two permutations in $\sigma_{k-1}H_{T_{k-1}}$, say $\pi \doteq
\sigma_{k-1} g$ and $\pi' \doteq \sigma_{k-1} g'$ for $g, g' \in
H_{T_{k-1}}$, belong to the same left coset of $\Gamma$ 
iff they belong to the same left coset of $\Gamma_{T_{k-1}}$. 
This follows because if $\sigma_{k-1} g' = \sigma_{k-1} g h$ for some
$h \in \Gamma$, then $h$ equals $g^{-1} g' \in H_{T_{k-1}}$, so $h \in \Gamma \cap
H_{T_{k-1}} = \Gamma_{T_{k-1}}$. 
\end{itemize}
\end{proof}

\begin{claim}\label{claim:2}
The count $c_v$ for $v \in [n]$ is nonzero iff $v$ is the minimum of
some $\Gamma_{T_{k-1}}$-orbit contained in the
$H_{T_{k-1}}$-orbit of $\sigma_{k-1}(k)$. 
\end{claim}

\begin{proof}
The set of values of $\pi(k)$ when $\pi$ ranges over $\sigma_{k-1}
H_{T_{k-1}}$ is the $H_{T_{k-1}}$-orbit of $\sigma_{k-1}(k)$. Since
$\Gamma_{T_{k-1}}$ is a subgroup of $H_{T_{k-1}}$, this orbit is the union
of some $\Gamma_{T_{k-1}}$-orbits. Combined with Claim~\ref{claim:1} and
the construction of the canonical representatives modulo
$\Gamma_{T_{k-1}}$, this implies Claim~\ref{claim:2}.
\end{proof}

\begin{claim}\label{claim:3}
If a count $c_v$ is nonzero then it equals
$|H_{T_{k-1}\cup\{v\}}|/|\Gamma_{T_{k-1}\cup\{v\}}|$. 
\end{claim}

\begin{proof}
Since the count is nonzero, there exists a permutation $\sigma' \in H$
that is a canonical representative modulo $\Gamma$ that agrees with
$\sigma_{k-1}$ on $[k-1]$ and satisfies $\sigma'(k)=v$. Applying 
Claim~\ref{claim:1} with $\sigma$ replaced by $\sigma'$, $k$ by $k'
\doteq k+1$, $T_{k-1}$ by $T'_k \doteq T_{k-1} \cup \{v\}$, and $\sigma_{k-1}$
by any permutation $\sigma'_k \in H$ that agrees with $\sigma'$ on
$[k]$, yields Claim~\ref{claim:3}. This is because the number of
canonical representatives modulo $\Gamma_{T'_k}$ in 
$\sigma'_kH_{T'_k}$ equals the number of (left) cosets of
$\Gamma_{T'_k}$ in $H_{T'_k}$, which is the quantity stated in
Claim~\ref{claim:3}.  
\end{proof}

The algorithm builds a sequence of permutations $\sigma_0,
\sigma_1, \ldots, \sigma_n \in H$ such that $\sigma_k$ agrees with
$\sigma$ on $[k]$. It starts with the identity permutation $\sigma_0 =
id$, builds $\sigma_k$ out of $\sigma_{k-1}$ for increasing values of
$k \in [n]$, and outputs the permutation $\sigma_n = \sigma$. 

Pseudocode for the algorithm is presented in Algorithm~\ref{alg:T}.
Note that the pseudocode modifies the arguments $\Gamma$, $H$, and $i$
along the way. Whenever a group is referenced in the pseudocode, the
actual reference is to a list of generators for that group.

\begin{algorithm}[t]
\caption{}\label{alg:T}
\begin{algorithmic}[1]
\Require positive integer $n$, $\Gamma \leq H \leq S_n$, $i \in [|H|/|\Gamma|]$
\Ensure lexicographically $i$-th canonical representative modulo $\Gamma$ in
$H$ 
\State $\sigma_0 \gets id$
\For{$k=1$ to $n$} 
  \State{$O_1, O_2, \ldots$ $\gets$ $\Gamma$-orbits contained in the
  $H$-orbit of $\sigma_{k-1}(k)$, in increasing order of
  $\min(O_i)$}
 \State{find integer $\ell$ such that $\sum_{j=1}^{\ell-1} c_{\min(O_j)} 
      < i \leq \sum_{j=1}^\ell c_{\min(O_j)}$, where $c_v \doteq
      |H_v| / |\Gamma_v|$}
  \State{$i \gets  i - \sum_{i=1}^{\ell-1} c_{\min(O_j)}$}
  \State{$m \gets \min(O_\ell)$}
  \State{find $\tau \in H$ such that $\tau(\sigma_{k-1}(k)) = m$}
  \State{$\sigma_k \gets \sigma_{k-1} \tau$}
  \State{$H \gets H_m$; $\Gamma \gets \Gamma_m$}
\EndFor
\State \textbf{return} $\sigma_n$
\end{algorithmic}
\end{algorithm}

The correctness of the algorithm follows from Claims~\ref{claim:2} and
\ref{claim:3}. The fact that the algorithm runs in polynomial time
follows from Proposition~\ref{prop:cgt}.
\end{proof}

\paragraph{Normal Form.}
Finally, we use the canonization captured in
Definition~\ref{def:canonical} and Lemma~\ref{lemma:canonical} to
establish the normal form for permutation groups given by
Lemma~\ref{lemma:permutation-group-normal-form} (restated below): 
\setcounter{lemma}{\value{normal-form}}
\begin{lemma} 
There is a polynomial-time algorithm $\nu$ that takes as input a list
$L$ of elements of $S_n$, and outputs a list of generators for the
subgroup generated by the elements in $L$ such that for any two lists
$L_0, L_1$ of elements of $S_n$ that generate the same subgroup,
$\nu(L_0) = \nu(L_1)$. 
\end{lemma}

\begin{proof}
Let $\Gamma$ denote the subgroup generated by $L$, and recall 
that $\Gamma_{[i]}$ denotes the subgroup of $\Gamma$ that
stabilizes each element in $[i]$, for $i \in \{0,1,\ldots,n\}$. We have
that $\Gamma_{[0]} = \Gamma$, and $\Gamma_{[n-1]}$ consists of the
identity only. 

We define $\nu(L)$ as follows. Start with $\nu$ being
the empty list. For $i \in [n-1]$, in the $i$-th step we consider each 
$j \in [n]$ that is in the $\Gamma_{[i-1]}$-orbit of $i$ in
order. Note that for each such $j$, the permutations in
$\Gamma_{[i-1]}$ that map $i$ to $j$ form a coset of 
$\Gamma_{[i-1]} \bmod \Gamma_{[i]}$. We append the
canonical representative of this coset to $\nu$. $\nu(L)$ is the value
of $\nu$ after step $n-1$.

As we only include permutations from $\Gamma$, $\nu(L)$ generates a  
subgroup of $\Gamma$. By construction, for each $i \in [n-1]$, 
the permutations we add in the $i$-th step represent all cosets of
$\Gamma_{[i-1]} \bmod \Gamma_{[i]}$. It follows by induction on $n-i$
that the permutations added to $\nu$ during and after the $i$-th step 
generate $\Gamma_{[i-1]}$ for $i \in [n]$. Thus, $\nu(L)$ generates
$\Gamma_{[0]} = \Gamma$. 

That $\nu(L)$ only depends on the subgroup $\Gamma$ generated by $L$
follows from its definition, which only refers to the abstract groups
$\Gamma_{[i]}$, their cosets, and their canonical
representatives. That $\nu(L)$ can be computed in polynomial time
follows by tracking a set of generators for the subgroups
$\Gamma_{[i]}$ based on Proposition~\ref{prop:cgt}. More specifically,
we use item 2 to check whether a given $j$ is in the
$\Gamma_{[i-1]}$-orbit of $i$, and item 3 to obtain $\Gamma_{[i]}$ out
of $\Gamma_{[i-1]}$ as $\Gamma_{[i]} = (\Gamma_{[i-1]})_i$. 
\end{proof}

\end{document}